\documentclass[fleqn,usenatbib,useAMS]{mnras}

\usepackage{graphicx}	
\usepackage{amsmath}	
\usepackage{multicol}        
\usepackage{bm}		
\usepackage{pdflscape}	
\usepackage{hyperref} 
\usepackage[version=4,arrows=pgf-filled, mathfontname=mathsf]{mhchem} 
\usepackage{booktabs} 

\newcommand{\maps}{MAPS} 
\newcommand{\psj}{PSJ} 

\usepackage[T1]{fontenc}
\usepackage{ae,aecompl}

\usepackage{newtxtext,newtxmath}

\title[Spatiotemporal coincidence of meteorites]{On the spatiotemporal coincidence of meteorites in recent fall search campaigns}

\author[Grèbol-Tomàs et al.]{P. Grèbol-Tomàs$^{1, 2}$%
\thanks{Contact e-mail: \href{mailto:grebol@ice.csic.es}{grebol@ice.csic.es}}, E. Peña-Asensio$^{3}$, J. M. Trigo-Rodríguez$^{1, 2}$, J. Ibáñez-Insa$^{4}$
\\
$^{1}$Institut de Ciències de l'Espai (ICE-CSIC), C/ Can Magrans s/n, Campus UAB, 08193 Cerdanyola del Vallès, Barcelona, Catalonia, Spain\\
$^{2}$Institut d'Estudis Espacials de Catalunya (IEEC), C/ Esteve Terradas 1, Campus Baix Llobregat - UPC, 08860 Castelldefels, Barcelona, Catalonia, Spain\\
$^{3}$Dipartimento di Scienze e Tecnologie Aerospaziali, Politecnico di Milano (PoliMi), Via La Masa 34, Milano, 20156, Lombardia, Italy\\
$^{4}$Geosciences Barcelona (GEO3BCN-CSIC), C/ Lluís Solé i Sabarís s/n, 08028 Barcelona, Catalonia, Spain}

\date{Last updated 2020 June 10; in original form 2013 September 5}

\pubyear{2024}

\begin{document}
\label{firstpage}
\pagerange{\pageref{firstpage}--\pageref{lastpage}}
\maketitle

\begin{abstract}
The meteoritical community widely assumes that the probability of finding two meteorites from different falls laying in close proximity is negligible. However, recent studies have suggested that spatiotemporal coincidences may be critical when associating a meteorite with a witnessed fall. In this work, we estimate the number of accumulated meteorites---those resulting from past falls---that are present in landing regions of new falls, while accounting for the effects of terrestrial weathering. We present a simple, fast-computing model to estimate such probability, validated with a Monte Carlo approach based on dark flight computations from real meteorite-dropping fireball data. Considering meteorite masses higher than 10~g, our results indicate that in regions with minimal weathering, like Antarctica, the probability of encountering a previous meteorite within a new fall strewn field may be as high as $75\%$. In environments with higher weathering rates, like countryside or urban regions, this probability decreases to $\lesssim 1\%$. When considering the 30~g Lake Frome 006 meteorite coincidence case, the probability of recovering a non-related meteorite with an age of 3.2~kyr from a 0.7~km$^2$ search area is 6.9\%. In the case of the 1~kg Ischgl meteorite, the probability of coincidence with another fresh meteorite of similar mass is 0.06\% assuming a large strewn field of 210~km$^2$. Applied to the Almahata Sitta case, our model predicts a 11.3\% of coincidence with a previous meteorite fall. Our results strongly suggest that isotopic dating is essential before associating any meteorite with a witnessed fall.
\end{abstract}





\section{Introduction} \label{sec: intro}

Meteorites are the tangible samples accounting for the early history of the Solar System. Their parent bodies, large meteoroids or asteroids, impact Earth at hypersonic velocities and start their ablation phase due to the friction with the air molecules \citep{Ceplecha1998, Silber2018AdSpR}, which produce bright meteors or fireballs. Monitoring extremely bright fireballs allows reconstructing their luminous paths and deceleration curves to decipher the events that may have produced meteorites. Recovering meteorites is, in fact, the main goal of fireball camera networks spread over the world, such as the European Fireball Network \citep[EN; ][]{Oberst94_EFN, Ceplecha1953_EFN}, the Cameras for Allsky Meteor Surveillance \citep[CAMS; ][]{Jenniskens2011Icar}, the Spanish Meteor Network \citep[SPMN; ][]{Trigo04_SPMN}, the Desert Fireball Network \citep[DFN; ][]{Bland04_DFN}, the Fireball Recovery and InterPlanetary Observation Network \citep[FRIPON; ][]{Colas2020AA}, or the Global Meteor Network \citep[GMN; ][]{Vida2021MNRAS}.

Using multiple observations, the atmospheric flight of the impactor can be modeled, allowing to gain insights into its original orbit, as well as its physical properties \citep[e.g.,][]{Collins05, Trigo05, Trigo06_Villalbeto, TrigoRodriguez15_Annama, Sansom19_trajectories, PenaAsensio21_FireTOC, Carbognani23}. Since the end of the last century, several meteorites have been recovered from their recorded falls. Until October 2024, a total of 55 meteorites have been identified with their orbit\footnote{\url{https://www.meteoriteorbits.info/}}.

Dynamical models can be applied to the observational data to estimate the terminal masses of meteoroids \citep{Gritsevich2008_mass, Sansom19_alphabeta, 2023MNRAS5205173P, 2024MNRAS533L92P}. For non-zero terminal masses, the dark flight trajectory of the meteoroid can be estimated to narrow meteorite search areas \citep{Ceplecha1987}. The estimation of the region where the meteorite may have landed, known as the strewn field, has been improved in recent years \citep[e.g.][]{Moilanen21_strewnfields, Towner2022_darkflight}. Combined with an enhancement in the resolution of surveillance camera images, recent Traspena \citep{Andrade23_Traspena} or Winchcombe \citep{McMullan24_Winchcombe} meteorites show a remarkable match between calculated results and the actual finding locations. 

Traditional (and most common) meteorite searching campaigns involve sweeping the area by foot thanks to a trained team of committed searchers. More sophisticated methods now involve the coupled use of drones and machine learning techniques to recognize the visual and/or thermal characteristics of meteorites, completing the search of the theoretical strewn fields in much less time \citep{Anderson20_ML, Hill23_ML}. However, these methods require a large dataset of detections to train the model in the actual background soil where meteorites are expected to be found. The model needs to be fed with a large number of images, tagged depending on whether they include a meteorite. To date, these applications have only proved successful in desertic areas due to this limitation \citep{Anderson22_ML}.

While meteorites remain in open air subjected to the changing weather conditions, several physicochemical processes can occur over the meteorite minerals, which degrade them \citep{BuchwaldClarke1989_weatheringreactions, Wlotzka93_weathering, AlKathiri05, Bland06_weathering}. This process is known as \textit{weathering}. Weathering acts as a time trial when looking for meteorites. The degradation causes meteorites to alter their mineral composition, ultimately mimicking terrestrial rocks. Meteorites are thus classified based on the \textit{degree of weathering}, which labels these effects \citep[W0-6,][]{AlKathiri05}. The degree of weathering is affected by several environmental factors, such as exposure to water, environmental and geological conditions or microbial activity. \citep{Gritsevich24_Ischgl}. Therefore, weathering is intrinsically dependent on the local environment where a meteorite lands and significantly influences the highest possible age of the meteorites that can be found in a region. For example, some Antarctic meteorites have been found to exhibit terrestrial ages higher than 100~kyr \citep{Nishiizumi95, Welten08_FRO01149, Welten06_antarcticages}, while in hot deserts meteorites have typical terrestrial ages lower than 20~kyr \citep{Gattacceca11, AlKathiri05}. In urban-close regions, terrestrial ages are expected to be much lower, but these are not characterized in the scientific literature. Moreover, meteorite sampling is related to the researcher's ability to identify these objects, e.g.,~smaller meteorites are far easier to identify in a white (icy) background than in a green forest. Environmental effects (if the meteorite has not been recovered) and proper storage and handling can change the degree of weathering of a meteorite over time.

During new fall searching campaigns, the presence of a background of previous non-related meteorites is often not considered. These are meteorites that were already in place when the observed fireball occurred, the latter called \textit{falls}. In this sense, the fact that two unrelated meteorites have fallen nearby is often not taken into account in calculations \citep{Benoit00_pairing, Hutzler16_pairing, Gallardo22_pairing}. Thus, if in a meteorite searching campaign two meteorites are found close to each other, they are initially associated with the same fall based on the pairing factor \citep{Hutzler16_pairing}. Once recovered, petrologic and isotopic affinities are used to confirm such pairings. In general, meteorites originating from the same fall are expected to exhibit similar residence times, as well as comparable chemical and petrological characteristics. Furthermore, if a sufficient number of fragments from a single fall are recovered, it becomes possible to infer the size of the original meteoroid’s parent body—assumed to be consistent across the fragments. This, in turn, facilitates the association of individual meteorites with a specific meteorite fall. However, some falls \citep[e.g., Almahata Sitta or Motopi Pan;][]{Goodrichetal_2019, Jenniskens21_MotopiPan} have been reported to show chemical and petrological differences among meteorites from the same fall. Whether these differences truly reflect distinct source bodies or can arise within a single fall may deserve further research. In such cases, additional tools are required to assess whether a recovered meteorite is genuinely related to a recent fall.

Measurements of Cosmic Ray Exposure Ages (CREA) are used to infer the residence time on the Earth's surface \citep{Leyaetal2003, Eugster2003, Eugsteretal2006}. The ratios of cosmogenic noble gases, particularly the ratio \ce{^22Ne}/\ce{^21Ne}, can be used to properly calculate the pre-atmospheric size of the meteoroid and the burial depth of the surviving meteorite within it. These factors are then used to compute the production rates and the respective CREA of the meteorite under study \citep{Wieleretal2016}.
    
In recent works, an estimation of the probability of coincidence among a fresh meteorite from a recorded fall and a \textit{find} (i.e. a fortuitous or unintentional discovered meteorite) has been demanded. In \citet{Gritsevich24_Ischgl}, the authors revisited the EN camera recordings, looking for a fireball that could explain the lucky find of the Ischgl meteorite. The best match was the EN241170 fireball, as the Ischgl meteorite was found inside its computed strewn field. However, the authors did not discard the possibility that the found meteorite might be related to a non-recorded previous fall. Moreover, \citet{Devillepoix22_minimoon} described that, despite finding a meteorite in the falling region predicted by the DFN after analyzing a meteorite-dropping fireball, its terrestrial age was $3.2 \pm 1.3$~kyr. Thus, this meteorite \citep[Lake Frome 006; ][]{Shober19} was indeed not related to the fireball recorded by the DFN. Their work estimated with fair assumptions that the upper limit of the probability of this coincidence to happen is around $2\%$.

Recent field evidence further underscores this issue. In two recent meteorite recovery campaigns in Oman (A. Zappatini, priv. comm.), search teams investigated: (i) the Al-Khadhaf fall, where 1.2 km$^2$ was searched on foot and two fresh meteorites \citep[8.21 g, 13.85 g;][]{MetBull2024_AlKhadhaf} were recovered, with no older meteorites identified; and (ii) the Raja fall, where 1.7 km$^2$ was searched, yielding one fresh \citep[26.81 g;][]{MetBull2025_Raja} and one older (3.70 g) meteorite. These examples demonstrate that systematic searches in hot desert environments can easily lead to the recovery of older meteorites within fresh fall areas.

The aim of the present work is to systematically assess the probability that two unrelated meteorites have fallen close by. For this purpose, we developed a model to estimate the number of accumulated meteorites already in place (Section~\ref{subsec: historical}). Subsequently, we evaluated the probability that, when looking for a fresh meteorite, a previous meteorite could be found in the same region. This issue was  tackled with an analytical approach based on the Poisson distribution (Section~\ref{subsec: poisson}) and with Monte Carlo simulations (Section~\ref{subsubsec: montecarlo}) based on actual meteorite landing regions from recent observational data. Our results enabled us to estimate such probability in a variety of environments and meteorite terrestrial ages (Section \ref{sec: results}). Finally, a summary of the results, conclusions, and remarks of this work are presented in Section~\ref{sec: conclusions}.

\section{Theoretical model and simulations}

Meteorites found during a search campaign can be categorized as either \textit{accumulated}--meteorites already present in the region before a witnessed fall-- or fresh. This classification relies on meteorite dating due to radioactive isotopes. Short-lived radionuclides (SLN), such as \ce{^7Be}, \ce{^{26}Al} or \ce{^{60}Co}, allow assessing the time of fall in a temporal resolution of years \citep{LeyaMasarick09, Povinec20, Rosen20}. On the other hand, long-lived radionuclides (LLN), such as \ce{^{14}C}, \ce{^{39}Cl} or \ce{^{41}Ca}, allow determining the time of falling with $\sim1000$ years of uncertainty \citep{Jull93_14Cdating_b, Hutzler16, Tauseef24}. Both SLN and LLN are formed from nuclear reactions of the solid body atoms with cosmic rays in outer space \citep{Alexeev19}. Once arrived to the Earth's surface, the meteorite is shielded from this radiation and the clock starts to count, as the live nuclides decay progressively.

Fresh meteorites incorporated SLNs during their exposure to outer space and, therefore, they still have a significant amount of them by the time they reach the ground. Any fresh fall should contain a considerable quantity of SLNs to prove its recent extraterrestrial history \citep{Llorca05_Villalbeto}. In contrast, accumulated meteorites do not contain SLNs but only feature a background of LLN. Generally, a meteorite is not considered a fresh meteorite anymore about 10~years after its fall. After this time, their SLNs have already decayed, for which it can not be distinguished from meteorites with higher residence times (M. Laubenstein; personal communication).

As meteorites fall in a single point, we modeled the fall of accumulated meteorites with a uniform distribution. Then, we simulated the regions where expedition teams would go searching for a fresh meteorite. We estimated the probability that, in that searching region, an accumulated meteorite was already there and could be found. This probability was assessed from a statistical point of view based on a Poisson distribution, which estimated it straightforwardly from simple inputs, and also using Monte Carlo simulations, which accounted for typical variations in the computed strewn fields and allowed validating the usefulness of the Poisson method.

\subsection{Number of accumulated meteorites} \label{subsec: historical}
We first determined the number of accumulated meteorites to be generated in our models. When a meteorite falls to the ground, weathering effects start taking place. They eventually become indistinguishable from surrounding terrestrial rocks, effectively removing them from the meteorite background. The exact timescale for this meteorite removal depends on the specific weather conditions of the place. According to the exponential decay model from \citet{Jull93_weathering}, the weathering effect obeys an exponential law with weathering constant $\lambda$, which decreases the amount of recoverable meteorites through time. 

We based our model on the meteorite accumulation rate equation proposed by \citet{Zolensky90_weathering}. Let the accumulation rate be defined as $\mathcal{R}$, in units of meteorites per km$^2$ per yr. Then, the total number density of preserved falls ($n$, in meteorites per km$^2$) changes with time ($t$) as:
\begin{equation}
    \frac{dn}{dt} = -\lambda n + \mathcal{R}.
    \label{eq: dndt}
\end{equation}
This equation accounts for both weathering meteorite removal and for meteorite addition rate. We assumed a constant weathering, independent of the meteorite mass or size. However, one may expect that weathering should not apply equally to all meteorite sizes. For example, weathering might be more important in smaller meteorites, where the surface-to-volume ratio is higher. As we did not find in the literature a relation between $\lambda$ and the meteorite mass, $m$, we considered it constant, that is, $\lambda (m) = \lambda$.

Meteorite addition as conceived by \citet{Zolensky90_weathering} can be interpreted as the incoming flux rate of meteorites, that is, $\mathcal{R}(m) = \mathcal{F}(m)$. With this, we integrated Equation \ref{eq: dndt} and rearranged the resulting terms to obtain the number density of accumulated meteorites ($n_a$) at a given time ($t$):
\begin{align}
        \int_{n_0}^{n_a} \frac{dn}{-\lambda n + \mathcal{F}} &= \int_0^{t} dt\nonumber\\
        \Longrightarrow n_a(m , t) &= \left(n_0 - \frac{\mathcal{F}(m)}{\lambda}\right) e^{-\lambda t} + \frac{\mathcal{F}(m)}{\lambda}.\label{eq: integral dndt}
\end{align}
Natural or anthropogenic causes (erosion, agricultural activities, floods, wildfires, warfare...) produce a potential reset of the number of meteorites in a region. Intending to include these effects in our model, we set the initial number density value to $n_0 = 0$. This assumption would also be valid for large meteorite fluxes for which the initial number of meteorites is negligible, that is, ${n_0 \ll \mathcal{F}(m)/ \lambda}$. Under this assumption, the number density of accumulated meteorites after a time $t_a$ after resetting is:
\begin{equation}
        n_a(m, t_a) = \frac{\mathcal{F}(m)}{\lambda} \left( 1- e^{-\lambda t_a}\right). \label{eq: n model}
\end{equation}
If $t_a\rightarrow+\infty$ then $n_a \rightarrow \mathcal{F}/\lambda$, which is the ratio between the falling rate and the weathering rate. This value also represents the maximum number density of meteorites that can be found in the region, as well as being the stationary state solution for Equation~\ref{eq: dndt}. In such case, the rate of incoming meteorites equals the rate of meteorites disappearing due to weathering. From a statistical point of view, $t_a$ might be interpreted as the terrestrial age of the oldest meteorite to be found in the working region.

In a general sense, one could also estimate the number density of accumulated meteorites that have fallen within a time interval $t~\in~\left[ t_1, t_2 \right]$ after resetting by integrating Equation \ref{eq: integral dndt} within the desired time span:
\begin{equation}
        n_a(m, t_1, t_2) = \frac{\mathcal{F}(m)}{\lambda} \left( e^{-\lambda t_1}- e^{-\lambda t_2}\right).
\end{equation}

In hot-desert regions, such as the Sahara or Atacama, mild meteorite removal may also be present. This physical removal can be conceptually included in $\lambda$, as meteorites would be picked-up from the ground. Indeed, \citet{Zolensky90_weathering} estimated $\lambda$ based on the actual number of meteorites found by its expedition team. If meteorite-hunters removal would be highly effective, that would reset the number of accumulated meteorites in the region, setting $n_0 = 0$, for which $n_a$ could be described as Equation~\ref{eq: n model} indistinctively.

\begin{figure}
    \centering
    \includegraphics[width=\columnwidth]{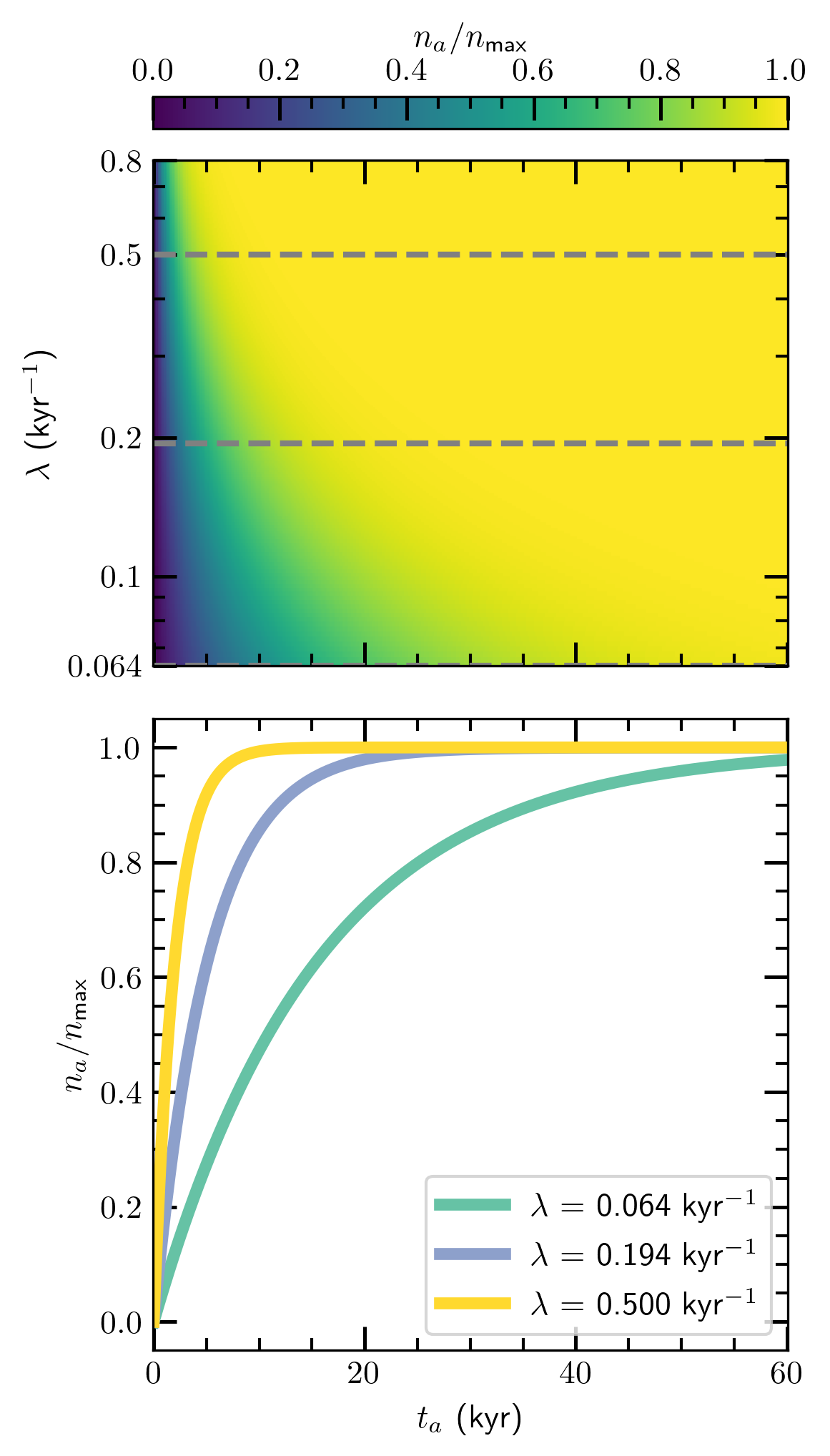}
    \caption{\textit{Top: }fraction of the maximum number of accumulated meteorites as a function of the weathering factor ($\lambda$) and the maximum meteoritic terrestrial age to be found ($t_a$), calculated using Equation~\ref{eq: n model}. \textit{Bottom: } number density of accumulated meteorites through time, normalized by its maximum value (i. e. cutout of the top plot). The stationary state is reached when $n_a/n_\text{max} \sim1$, with $n_\text{max} = \mathcal{F}{/\lambda}$.}
    \label{fig: nnmax}
\end{figure}

Figure~\ref{fig: nnmax} shows the resulting dependence of $n_a$ as stated in Equation~\ref{eq: n model}. In both plots, the number density of meteorites over time was normalized by the maximum number density of meteorites, being ${n_\text{max} = \mathcal{F}/\lambda}$. From the top plot it can be read which is the transient time required to reach the stationary state, that is, when ${n_a/n_\text{max} \sim 1}$ and, consequently, the number of accumulated meteorites remains constant through time. This minimum time depends on the  weathering factor, $\lambda$. 

The bottom plot in Figure \ref{fig: nnmax} represents a cut of the top plot at fixed $\lambda$ values, clearly showing two expected regimes. At low $t_a$, after resetting or removal of accumulated meteorites in the region, there is a relatively fast increase in the number of existing meteorites. In the long-term regime, if no further reset has taken place, the system evolves to the stationary state.  The value $\lambda = 0.064$~kyr$^{-1}$ was measured for Antarctic meteorites by \citet{Bland96}, while $\lambda = 0.197$~kyr$^{-1}$ was calculated with Roosevelt County meteorites by \citet{Zolensky90_weathering} with \citet{Boeckl72_terrestrialage} data. These are the only two references in literature reporting $\lambda$ values. The value of $\lambda = 0.5$~kyr$^{-1}$ was added for representation purposes. The probability that two or more meteorites have fallen close by was further evaluated in this work as a function of both $\lambda$ and $t_a$.

In \citet{Evatt20_flux}, it is calculated the flux of meteorites reaching Earth as a function of its mass from Antarctic meteorite data. In their work, the authors provided the equatorial flux and modeled its variation with latitude by rescaling it with the latitudinal factor $\gamma (\theta)$ \citep{Evatt20_flux}. The authors also mention the average pairing factor, which represents how many meteorites are generated by each fall. For simplicity, we considered that each fall generated only one single meteorite. If this factor was considered, each meteoroid would fragment in several meteorites of lower mass, adding a hardly-modeling variable to our calculations (the mass of each fragment). Note that this would have incremented the number of accumulated meteorites fallen in an area. However, we might expect them to be clustered for coming from relatable falls, with low scattering on the ground. Hence, even though a higher number of accumulated meteorites may increase the probability of coincidence, we might expect their packaged distribution on the ground would strongly mitigate these increasing effects. Because fragmentation modeling of accumulated meteorites is not yet available, the probabilities reported here might be regarded as lower limits.

\begin{figure}
    \centering
    \includegraphics[width=1\columnwidth]{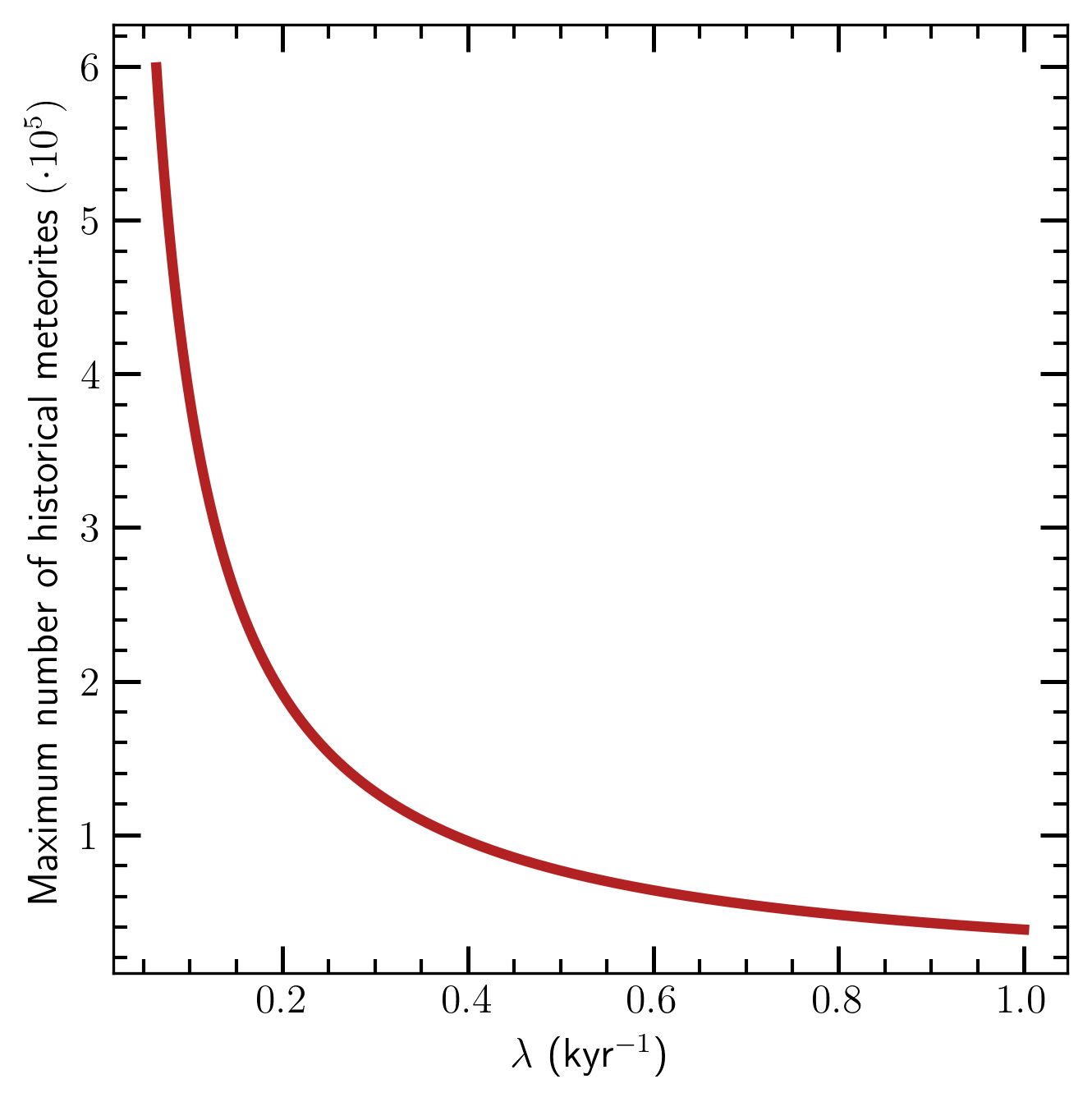}
    \caption{Maximum number of meteorites with mass higher than 10~g to be found in an area of ${A \sim 700,000}$~km$^2$ based on $\lambda$, according to Equation~\ref{eq: number meteorites}.}
    \label{fig: nmax}
\end{figure}

Overall, the final flux estimation used in this work, $\mathcal{F}(m)$, is given from the flux from \citet{Evatt20_flux}, $\mathfrak{F}(m)$\footnote{The flux given in \citet{Evatt20_flux} is a cumulative flux. As such, the value $\mathfrak{F}(m = M)$ with $M$ an arbitrary mass reads as `the equatorial flux of meteorites with a mass higher than $M$ reaching Earth'.}, as
\begin{equation}
    \mathcal{F}(m) = \mathfrak{F}(m) \cdot \gamma ( \theta).
    \label{eq: flux conversion}
\end{equation}
The expression of \citet{Evatt20_flux} for $\mathfrak{F}(m)$ represents the cumulative flux of meteorites with mass higher than $m$. According to their work, the flux at the equator for masses higher than 10~g is $\mathfrak{F}(m=10\text{~g}) \simeq 68.39$~km$^{-2}$~Myr$^{-1}$. In an area $A$, the number of accumulated meteorites in place after a time $t_a$ with a mass higher than $m = 10$~g can be calculated from Equation~\ref{eq: n model} as
\begin{align}
    N_a (m > 10\text{ g}, t_a) &= n_a(m > 10\text{ g}, t_a) \cdot A \nonumber\\
    &= \frac{\mathcal{F}(m = 10\text{ g}) \cdot A}{\lambda} \left( 1 -e^{-\lambda t_a} \right),
    \label{eq: number meteorites}
\end{align}
where $\mathcal{F}(m)$ is found according to Equation \ref{eq: flux conversion}. $N_a$ is the number that was used for the subsequent modeling of the probability that two meteorites have fallen close by (Section~\ref{subsec: models}).
As in Equation~\ref{eq: n model}, the maximum number of meteorites with mass higher than $m$ was calculated as
\begin{equation}
    N_\text{max} = \frac{\mathcal{F}(m) \cdot A}{\lambda},
    \label{eq: nmax}
\end{equation}
which is the number of meteorites on the ground when the stationary state is reached. Figure \ref{fig: nmax} shows the maximum number of meteorites with mass higher than 10~g to be found in an area of $A \sim700,000$~km$^2$, which would correspond to a region similar to Central Europe, the Great Victoria Desert (Australia), or Sumatra Island (Indonesia).

\subsection{Estimation of strewn-field distributions} \label{subsec: fresh}
In the case of a witnessed fall, a typical procedure for a research team involves analyzing camera images, computing the fireball's trajectory, and modeling its dark flight. The results of the dark flight modeling help identify the region where the team will search for the meteorite. In this study, we aimed to reproduce this standard approach, computing the landing regions of several meteorites as if they were calculated by research teams.

We modeled the fall of a fresh meteorite with a modified version of the DFN dark flight code \citep{Towner2022_darkflight}, in which dynamical equations are solved to find the landing point of a particle. The input parameters of the code were the initial latitude, longitude, height, velocity, slope, azimuth, mass, and density of the meteoroid. In order to estimate a fall region, we simulated 100 clones per fall, each one with different initial parameters extracted from a normal distribution. For a given fall, the combination of all the landing points on the surface results in the landing region where an expedition team would go to search for the meteorite. To simplify the problem, we fitted this landing region into an ellipse. The length of each ellipse axes was chosen from the normal distribution of the points along it, such that included $2\sigma$ (95.4\%) of the points.

Strewn fields were simulated based on real fireball data. We considered the falls reported by the EN recorded during the 2017-2018 period \citep{Borovicka2022_EFNdataI}. This unbiased fireball database includes the bright flight parameters for 824 events. From these, only 25 had a terminal mass higher than 10~g. This threshold is chosen according to the accumulated flux used in Section \ref{subsec: historical}. We took the nominal values of these events in the EN database as initial parameters to our dark flight simulation, except for the initial position and density. 

\begin{table*}
\centering
\caption{Errors for the initial parameters of the dark flight estimation. Values extracted from \citet{Borovicka2022_EFNdataI} according to the errors of their measurements, written in their main text. The columns represent the following properties, in order: latitude, longitude, azimuth, zenith distance of the meteoroid trajectory, terminal mass, terminal/maximum brightness height, and terminal velocity. No uncertainty was considered in the terminal mass trying to model the possible landing sites of a single meteoroid without fragmentation, as this process is represented by the 10~g mass simulations.}
\label{tab: errors}
\begin{tabular}{lccccccc}
\hline
\textbf{Parameter} & Lat. ($^\circ$) & Lon. ($^\circ$) & Azi. ($^\circ$) & Zen. dist. ($^\circ$) & T. mass (kg) & T./Max. bright. height (m) & T. vel (m/s) \\ \hline
\textbf{Error}     & 0.0001         & 0.0001           & 0.29           & 0.04                  & 0           & 20            & 263  \\
\hline
\end{tabular}
\end{table*}

The EN coverage spans an area of approximately $7 \times 10^5$~km$^2$, corresponding to Central Europe \citep{Borovicka2022_EFNdataI}. In our simulation, we assumed a rectangle working region defined by the vertexes coordinates (8.745858$^\circ$~E, 47.447117$^\circ$~N) and (23.552000$^\circ$~E, 53.4607833$^\circ$~N), which corresponds to an area of 700,000.003~km$^2$ and a latitudinal factor of $\gamma (\bar{\theta}) = 0.801$, where $\bar{\theta}$ is the mean latitude (Equation \ref{eq: flux conversion}). According to the flux estimation from \citet{Evatt20_flux} and the considered working area, the number of fresh meteorite falls with a mass higher than 10~g to be expected in the region within a 10 year span (maximum age to distinguish SLNs in fresh falls) is 392. This is the number of ellipses to be simulated with the present methodology.

The dark flights were initiated at random coordinates extracted from a random uniform distribution within the working area. Their azimuths were also taken from a random uniform distribution. The density of the meteoroids was assumed to be chondritic, extracted from a uniform distribution between 2000~kg~m$^{-3}$ and 3700~kg~m$^{-3}$ \citep{Ceplecha1998, Hilton02_densities, Collins05}. Finally, the shape parameter, determining the aerodynamic resistance while flying, was changed for each clone. This was also chosen from a random uniform distribution, between 1.1 and 2.7 \citep{MillerBailey79_shape,Zhdan07_shape,Gritsevich08_shape}. The drag coefficient for each clone mainly constributes to the length of its flight, and has direct impact on the extent of the strewn field.

The errors associated with the normal distribution to create each clone were the errors specified in \citet{Borovicka2022_EFNdataI}, reproduced in Table \ref{tab: errors}. In the case of latitude and longitude, the error used was the precision of the reported values. For heights, the highest value of the range of errors given in the reference was used. While height errors are typically larger, we kept a value of 20~m as it is the only value reported in the reference. Additionally, from some EN event data it can be seen that the poiny-by-point velocity dispersion is similar at the beginning and at the end of the flight \citep[e.g.][]{PenaAsensioGritsevich25}. Since the terminal velocity error is not reported, we consider the initial velocity errors as a proxy. As such, we used the 95\% percentile of the errors reported for the initial velocity for the terminal velocity error. Finally, in order to estimate the errors for azimuth and zenith distance (not reported in \citet{Borovicka2022_EFNdataI}), we converted the geocentric coordinates of the trajectory to local coordinates and then estimated the errors from the obtained azimuth and elevation.

The atmospheric density at each height during the meteoroid descent was computed using the pressure and temperatures extracted from a hydrostatic and linear models, respectively. The relative humidity profile was extracted from a semiempirical model based on data from \citet{PeixotoOort1992_book}. No wind profile was applied to avoid preferential directions in the resulting ellipses. 

For each of the 392 falls, we estimated their strewn fields by considering different initial heights, initial velocities and nominal masses. These are input parameters to the dark flight estimation code and critically influence the resulting strewn fields. This methodology was designed to explore the full range of plausible scenarios. The specific parameter variations applied to each fall are as follows:
\begin{itemize}
    \item Initial height: meteoroids were considered to start the dark flight at the terminal height but also from the height at maximum brightness. With the latter we intend to represent a higher altitude fragmentation, producing an early dark flight.
    \item Initial velocity: similar to the approach taken with initial height, we have considered scenarios in which the meteoroid’s velocity corresponds to its terminal velocity, as well as cases where the velocity is 3 km/s---the lower threshold at which ablation ceases and dark flight commences. \citep{Ceplecha1998, Moilanen21_strewnfields, Vida23}.
    \item Meteoroid mass: we considered that the meteoroid mass could be either the terminal mass or the minimum mass of 10~g, representing the range from the intact meteoroid to the smallest recoverable fragments. This variation served to approximate fragmentation effects.
\end{itemize}

Therefore, each fall was modeled under 8 distinct parameter combinations, resulting in 8 strewn fields per event. Additionally, for a fixed height and velocity, we generated composite ellipses spanning the full range of considered masses. This approach is aimed to capture the extreme bounds of possible strewn field geometries. In practice, the actual strewn field is expected to fall between the smallest and largest simulated ellipses. For instance, the most extensive fields typically resulted from simulations starting at the height of maximum brightness, rather than at terminal height (see Table~\ref{tab: areas summary}, discussed later).

\begin{figure}
    \centering
    \includegraphics[width=\columnwidth]{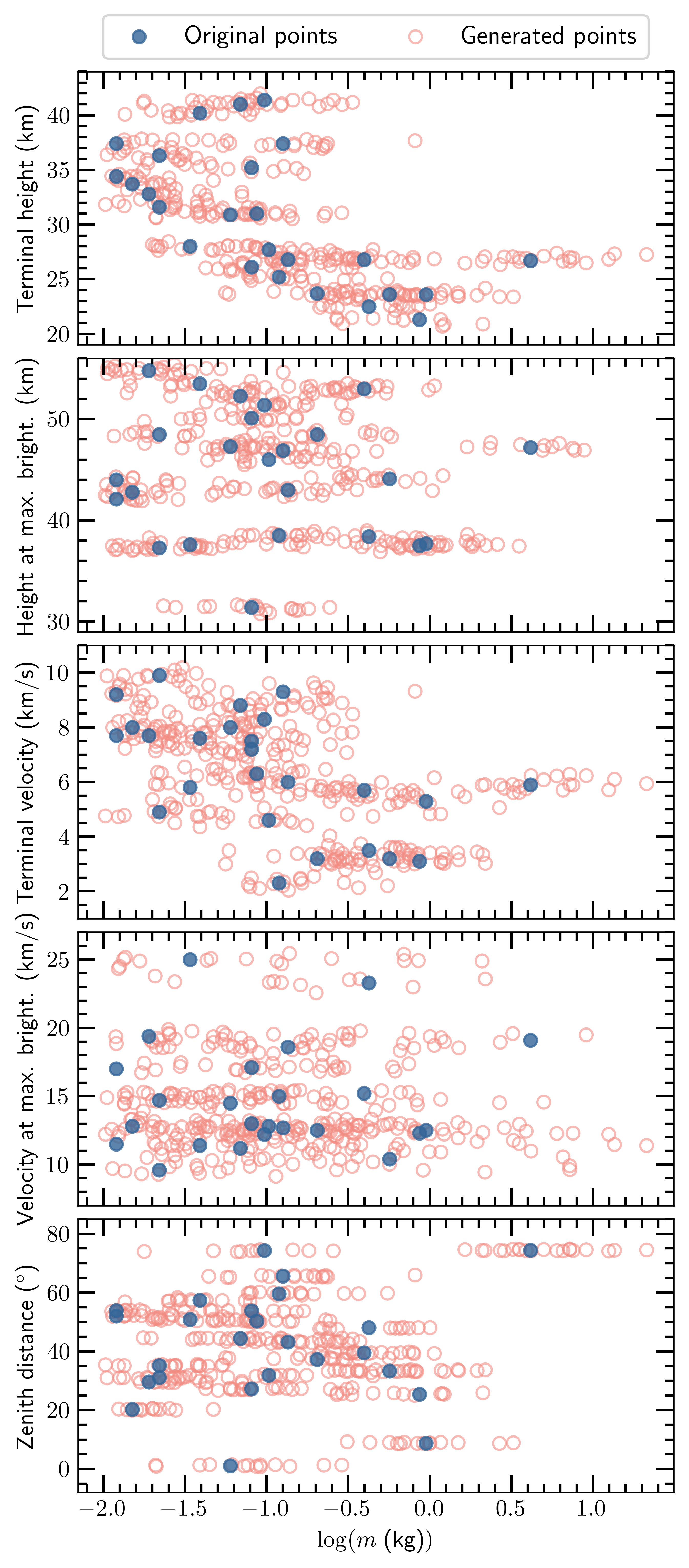}
    \caption{Nominal values of terminal height, height at maximum brightness, terminal velocity, velocity at maximum brightness, zenith distance, and mass from the recorded events in \citet{Borovicka2022_EFNdataI} with terminal mass higher than 10 g (original points), along with 392 events parameters generated to simulate fresh meteorites. Generated points were extracted from a multivariate probability distribution function accounting for the five parameters.}
    \label{fig: multivariate distribution}
\end{figure}

The random nominal values of mass, zenith distance, velocity and height were chosen from a multivariate random distribution \citep{Scott15_book, Bashtannyk01, Silverman98}, which was taken to reproduce the 25 suitable values in the EN database. The advantage of this procedure is that it correlates the behavior of several variables. The distribution estimation was performed with the Python's \texttt{scikit-learn} package \citep{pedregosa2011scikit} Gaussian kernel density estimator with bandwidth equal to 0.26. Figure \ref{fig: multivariate distribution} shows the values of the original 25~points from the database and the parameters of 392 events created with the multivariate PDF.

The ground level for the simulations was set to 440~m above the WGS84 ellipsoid \citep{WGS84_2014} by default, which is the mean height over the working area of the EN. For completeness, we repeated the simulations of the same falls setting the ground levels to 40~m and 840~m, but no clear difference was seen (Table~\ref{tab: areas summary}). We consider this is due to the different random effects taking place when simulating the 392 ellipses, each with different initial conditions.

\begin{table}
\caption{Summary of the characteristic areas obtained with the different combinations of initial heights, velocities, masses and ground levels. The keywords of height, velocity and mass include: value at maximum brightness (bri), value at terminal point (ter), minimum value (min; 3~km/s for velocity and 10~g for mass) and both possible cases (all). The characteristic areas of the distributions are the mode of the distribution of the strewn field areas within the 1.5~IQR.}
\label{tab: areas summary}
\resizebox{\columnwidth}{!}{
\begin{tabular}{lcccc}
\toprule
\textbf{Ground level} & \textbf{Height} & \textbf{Velocity} & \textbf{Mass} & \textbf{mode(}$\boldsymbol{A_{sf}}$\textbf{)}\\
\textbf{(m)} & & & & \textbf{(km)} \\ \midrule
40 & bri & min & min & 1.19 \\
40 & bri & min & ter & 1.62 \\
40 & ter & ter & min & 0.29 \\
40 & ter & ter & ter & 0.59 \\
40 & ter & min & ter & 0.54 \\
40 & ter & min & min & 0.27 \\
40 & bri & min & all & 1.69 \\
40 & ter & ter & all & 0.57 \\
40 & ter & min & all & 0.54 \\
40 & all & all & all & 2.15 \\
440 & bri & min & min & 1.23 \\
440 & bri & min & ter & 1.62 \\
440 & ter & ter & min & 0.28 \\
440 & ter & ter & ter & 0.59 \\
440 & ter & min & ter & 0.55 \\
440 & ter & min & min & 0.25 \\
440 & bri & min & all & 1.81 \\
440 & ter & ter & all & 0.58 \\
440 & ter & min & all & 0.54 \\
440 & all & all & all & 2.13 \\
840 & bri & min & min & 1.26 \\
840 & bri & min & ter & 1.62 \\
840 & ter & ter & min & 0.29 \\
840 & ter & ter & ter & 0.60 \\
840 & ter & min & ter & 0.53 \\
840 & ter & min & min & 0.28 \\
840 & bri & min & all & 1.69 \\
840 & ter & ter & all & 0.58 \\
840 & ter & min & all & 0.53 \\
840 & all & all & all & 2.02 \\ \bottomrule
\end{tabular}
}
\end{table}

\begin{figure*}
\includegraphics[width=1\textwidth]{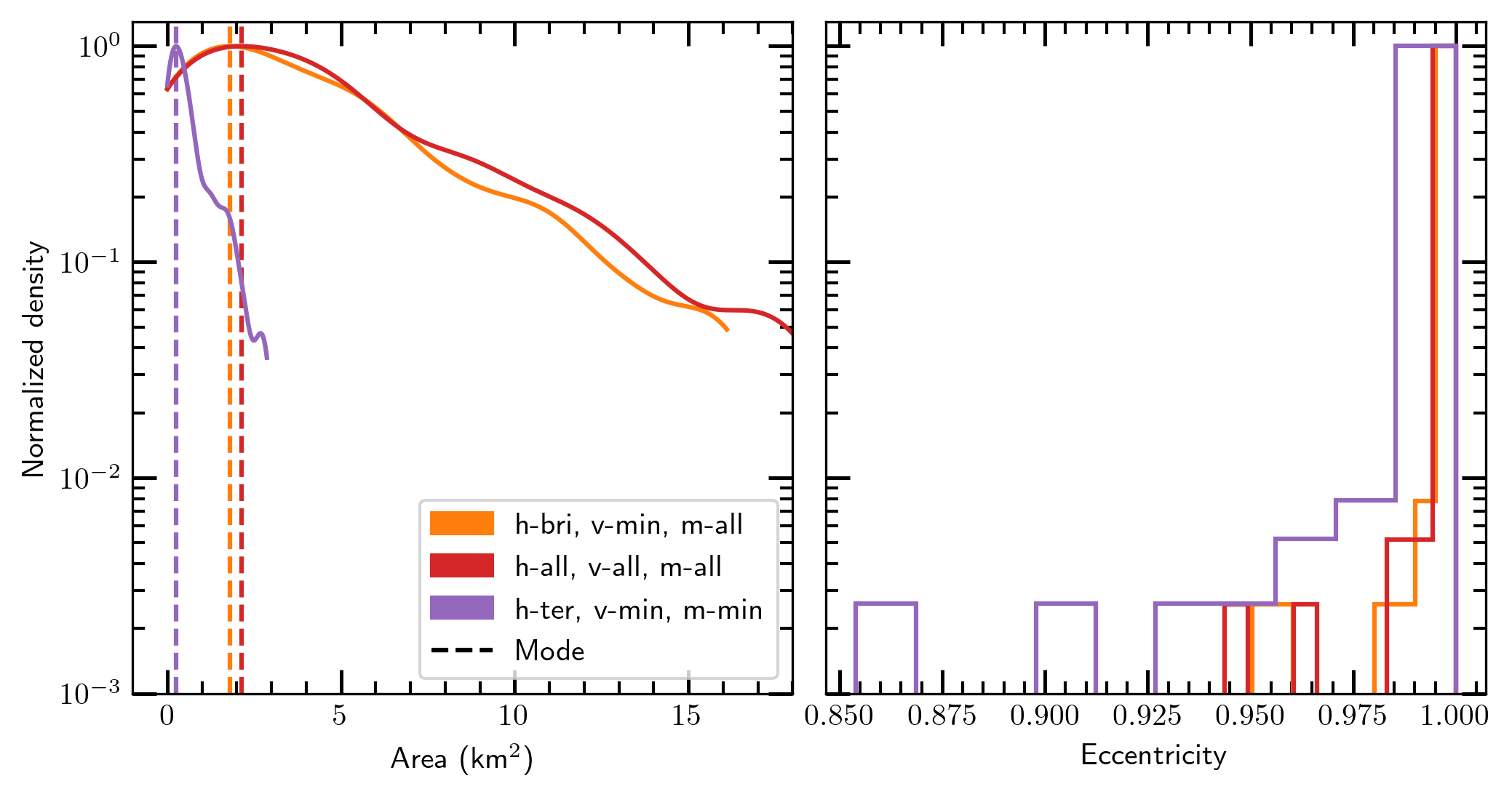}
\caption{Areas and eccentricity values of the 392 fresh meteorite ellipses, for different representative combinations of initial height, initial velocity and mass, at a ground level of 440~m. Legend follows the nomenclature of Table~\ref{tab: areas summary}. The plots have been normalized so that their respective maximum value is 1. \textit{Left: } kernel density estimator of the 1.5 IQR areas distribution. The vertical dashed lines correspond to the mode of the distribution (Table~\ref{tab: areas summary}). \textit{Right: } eccentricities distribution. The bin size was estimated with the Sturges method.}
\label{fig: ellipses data}
\end{figure*}

Figure~\ref{fig: ellipses data} shows the areas and eccentricities of the 392 ellipses generated, for three representative combinations of height, velocity and mass. Ellipses were characterized for being highly elongated, most of them with eccentricities higher than 0.98. This was due to the low errors in the parameters, except for the velocity (Table~\ref{tab: errors}). All clones for a given fall had similar initial coordinates, thus the difference between landing points lied along the azimuthal axis, where the velocity was being changed. Overall, the uncertainty in azimuth had a lower effect on the scattering of the landing points, compared to the velocity uncertainty. For this reason, the computed ellipses were highly eccentric.

The areas distributions were marked for having an important positive skewness and a long tail at large areas. Ellipses were characterized with the mode of the distribution, estimated with a gaussian kernel density estimator from the areas within the 1.5 interquartile range (IQR). This outlier filter was applied to prevent large, underrepresented areas from skewing the area estimates toward misleading high values. The resulting distribution is represented in Figure~\ref{fig: ellipses data}. Table~\ref{tab: areas summary} shows as summary of all the mode values, for the different combinations of initial height, velocity and mass, and ground level. The values are consistent with other strewn field area estimations \citep{Towner2022_darkflight, Devillepoix22_maduracave, GrebolTomas24_Portugal}.

As expected, the area values are higher if the dark flight starts at higher initial heights. That is, the area estimations for a maximum brightness height are higher than their terminal height counterparts. Dark flights starting at the maximum brightness height can generate ellipses doubling the area of those starting at terminal heights, which are 15 km lower in average. As previously noted, actual dark flights are expected to initiate at intermediate altitudes between these two heights, so the resulting strewn field area will lie between their respective values. Additionally, starting with the minimum velocity of 3~km/s makes the meteoroids to be strongly influenced by atmospheric drag due to their low kinetic energy. As such, their dark flights cover smaller distances than faster meteoroids. Finally, atmospheric drag is also more present in meteoroids of lower mass, having similar effects than a low initial velocity. Should a wind profile had been applied, we may expect the resulting ellipses tending to be biased in the direction of the wind. For light winds, we may expect the total area of highest density to be the same, but with a shape different from an ellipse. The reader is invited to use the strewn field area estimation that best fits their needs.

Finally, we performed an analysis to evaluate how the strewn field area would change if the ground level was varied from a few meters to 2~km. Our results indicated that there were no substantial changes in the area.

We considered the case including all simulated points (for all initial heights, velocities and masses) to be the most similar to an actual dark flight modeling that research teams would conduct. As in the other cases, these ellipses were estimated so that they included 2$\sigma$ of all the points (800 points per ellipse). Its mode was later used in this work as a representative ellipse area. 

\begin{figure}
    \centering
    \includegraphics[width=1\columnwidth]{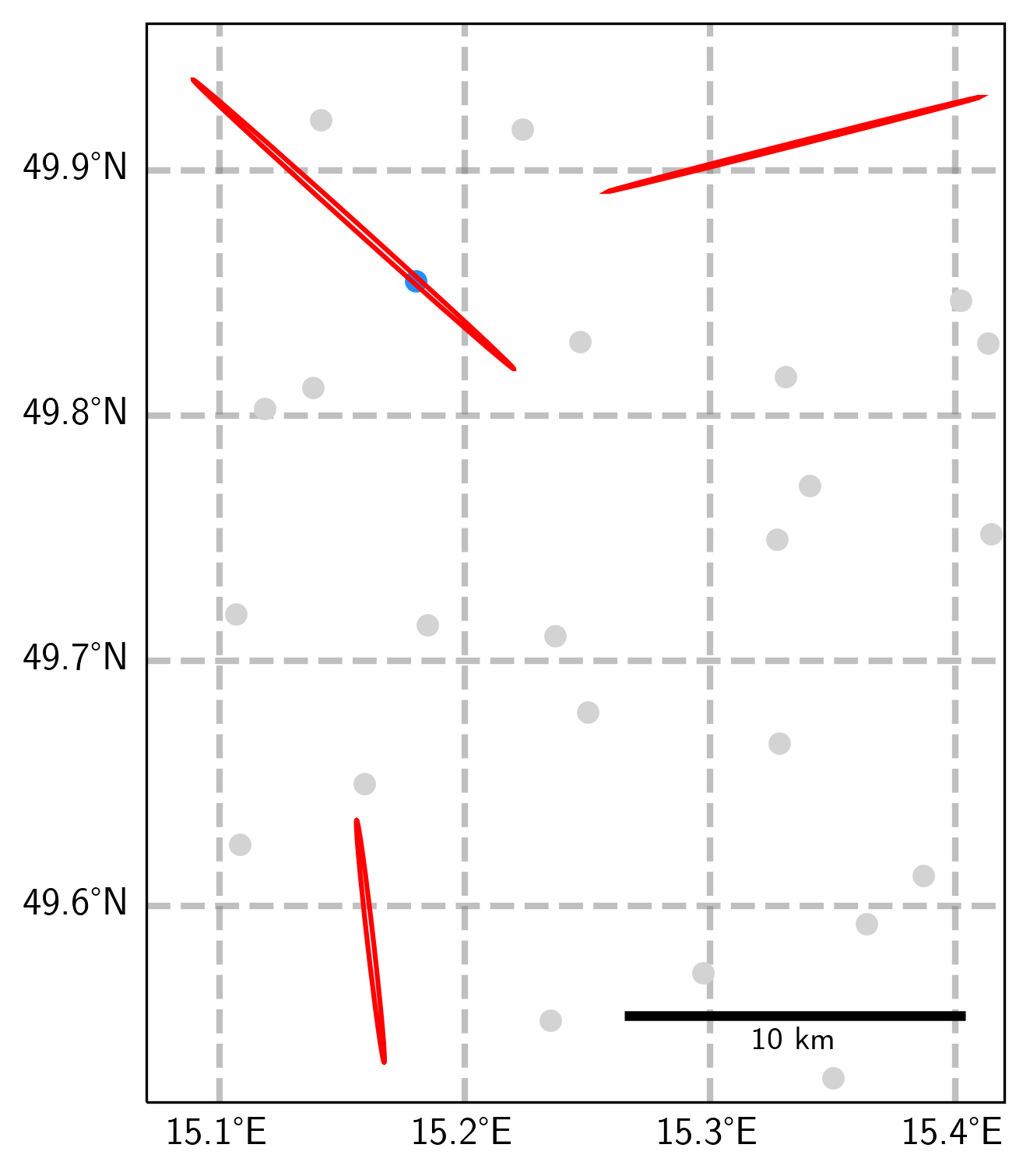}
    \caption{Zoom-in to a study-case scenario, corresponding to an approximate area of $\sim1,200$~km$^2$ in Central Europe. Accumulated meteorites were randomly distributed in the region and represented with grey dots. Three ellipses in the considered area are seen in red, corresponding to the regions where a research team would look for fresh meteorites. There is one coincidence between a fresh meteorite and an accumulated meteorite in the upper-left ellipse, colored as blue. Scale bar is 10~km.}
    \label{fig: on land}
\end{figure}

 As a matter of example, Figure \ref{fig: on land} shows a zoom-in to the simulated Central European region. Accumulated meteorites are randomly spread through the region, and fresh meteorite ellipses are shown. In this particular image, one of the ellipses appears to include one accumulated meteorite inside. This accounted for a coincidence in the probability estimations (Section \ref{subsec: results simulations}).

\subsection{Probability of two accumulated meteorites falling nearby} \label{subsubsec: haversine}
For completeness, we evaluated the probability that two accumulated meteorites have fallen nearby. In some meteorite search campaigns, meteorite pairing is still an issue \citep{Gallardo22_pairing}. It is often hard to distinguish whether two close meteorites come from the same fall. Using the maximum number of meteorites shown in Figure~\ref{fig: nmax}, we made an estimation as a function of $\lambda$ of the number of meteorites from a different fall expected to be recovered within a given area. 

For two points at latitude-longitude coordinates $\left( \phi_i, \theta_i\right)$ their distance $r$ on the surface of a sphere is given by the haversine formula:
\begin{equation}
    \text{hav}\left(\frac{r}{R}\right) = \text{hav}\left(\phi_1 - \phi_0\right) + \cos\left(\phi_0\right)\cos\left(\phi_1\right)\text{hav}\left(\theta_1 - \theta_0 \right),
    \label{eq: haversine}
\end{equation}
where $R$ is the Earth radius and $\text{hav}(z)=\sin^2\left(z/2\right)$ is the haversine of $z$. This expression has been proven as a fast and accurate method to calculate distances on Earth's surface under the assumption of its spherical shape \citep{Sinnott1984_haversine, Gade_2010_haversine}. We computed the probability density function (PDF) of the distance between two meteorites within the selected area, the results of which are shown in Appendix~\ref{app: haversine pdf}. The values of the PDF describe how likely it is for two meteorites falling at random positions to be separated a given distance. These values were later used in the estimations of Section~\ref{subsec: number density historical}. Note that the PDF distribution does depend on the shape of the working region.

\subsection{Calculating meteorite coincidence probability}\label{subsec: models}
\subsubsection{The Poisson distribution method approach}
\label{subsec: poisson}
The probability of having accumulated meteorites inside a fresh meteorite falling area (i.e., the area with the highest likelihood to recover a sample defined by the strewn field) can be addressed from the point of view of analytical statistics. Meteorite fall can be assumed to be a discrete process, with no spatial or temporal correlation between falls. As such, meteorite fall occurrence can be represented by a Poisson distribution. In its general form, the expected probability of $k$ events occurring with a constant rate of $\nu$ events per interval is given by:
\begin{equation}
    P(X = k) = \frac{\nu^k e^{-\nu}}{k!}.
    \label{eq: Poisson definition}
\end{equation}
The value $\nu$ represents the expected number of meteorites in a given area. The particular $P(X = 1)$ case represents the probability of finding only one accumulated meteorite inside the falling area of fresh meteorite. Assuming a strewn field area $A_{sf}$ for fresh meteorites, the expected number of accumulated meteorites within that area ($N_{a,\;sf}$) is given by Equation~\ref{eq: number meteorites}, changing the parameter $A$ accordingly. It is worth noticing that Poisson modeling does not depend on the shape of the strewn field, but only on its area. 

Therefore, the probability of finding one or more meteorites (i.e., with no restriction on the strewn field occupation number) is
\begin{equation}
    P (X \geq 1) = 1 - P(X=0) = 1-e^{ -N_{a,\;sf}},
    \label{eq: Poisson Nh}
\end{equation}
with $N_{a,\;sf}$ given by Equation \ref{eq: number meteorites} for a fixed $(\lambda, t_a)$ pair and the characteristic area of the strewn fields. Overall, Equation~\ref{eq: Poisson Nh} provides a straightforward method for estimating the probability of coincidence, requiring only basic input parameters.

\subsubsection{The Monte Carlo simulations approach} \label{subsubsec: montecarlo}

Monte Carlo simulations were used to validate the Poisson distribution method (Section~\ref{subsec: poisson}) by using an alternative approach to assess the probability of meteorite spatiotemporal coincidence.

Estimating the probability that an accumulated meteorite can be found within the strewn field of a fresh meteorite using simulations involved the 392 ellipses computed in Section \ref{subsec: fresh}, taking the flux latitudinal factor correction of $\gamma (\bar{\theta}) = 0.801$. This latitudinal factor is that of the working rectangle defined in Section~\ref{subsec: fresh}, in the same region where the EN operated. The probability of coincidence was calculated by dividing the number of occupied ellipses by the total number of simulated ellipses. A total of 1000 scenarios for a given $(\lambda , t_a)$ pair were computationally generated. In each scenario, the 392 computed ellipses were fixed in position, but the position of accumulated meteorites was randomly changed. Thus, the total number of ellipses from which to perform statistics was $3.92 \cdot 10^5$. For each $(\lambda , t_a)$ pair, the number of accumulated meteorites, $N_a$, was constant throughout the rectangle working region (Section \ref{subsec: fresh} and Figure~\ref{fig: nnmax}). Accumulated meteorite positions were extracted from a uniform distribution on the surface of a sphere \citep[Appendix \ref{app: haversine pdf};][]{Feller1971_ProbTheory}. 

\section{Results and discussion} \label{sec: results}
As mentioned in Section \ref{subsec: historical}, the potential number of accumulated meteorites in a given area (Equation~\ref{eq: number meteorites}) is deeply related to the values $\lambda$ and $t_a$, which are generally unknown. In fact, the probability of coincidence that an accumulated meteorite could be found inside a fresh meteorite search area is expected to depend on $N_a$. Thus, in Section~\ref{subsec: number density historical} we modeled the number density of accumulated meteorites to be found within a given area. For completeness, we developed a simple approach to estimate the weathering constant $\lambda$ of a non-desertic region based on the known recovered meteorites (Section~\ref{subsec: lambda in urban regions}). This provides us with a lower limit of the $\lambda$ values to be expected in non-desertic regions.

Throughout this section, we used the meteorite flux estimated from Antarctic data from \citet{Evatt20_flux}. For full disclosure, in Appendix~\ref{app: mass distribution comparison} we compared it with the mass distribution from the EN database and concluded that the flux of meteorites has not substantially changed over the last kiloyears. Therefore, Equation~\ref{eq: flux conversion} was suitable to describe the current meteorite fall flux.

The probability of coincidence was assessed from an analytical point of view (Section~\ref{subsec: results statistical analysis}) and a Monte Carlo approach (Section~\ref{subsec: results simulations}). The analytical method was based on a simple modeling based on the Poisson distribution (Section~\ref{subsec: poisson}).

Finally, we applied our probability estimations to real cases: the Lake Frome 006, Ischgl and Almahata Sitta meteorites (Section~\ref{subsec: application to real cases}). Except for the application to real cases, our probability estimations considered the coincidences between meteorites of masses higher than 10~g and the EN flux latitudinal factor correction, $\gamma = 0.801$.

\subsection{Number density of accumulated meteorites} \label{subsec: number density historical}
The probability of coincidence in a given area is determined by the number of accumulated meteorites. Thus, it is first necessary to characterize the number density of accumulated meteorites within a region, which is expected to mainly depend on the specific environmental conditions of the place. As already mentioned, these are taken into account through the weathering factor $\lambda$.

Assuming that the accumulated meteorites are randomly distributed in the rectangular working area of Section~\ref{subsec: fresh} (${A \sim700,000}$~km$^2$), we computationally estimated the PDF of the distance between two meteorites, $r$, given by the haversine function (Equation~\ref{eq: haversine}, Appendix \ref{app: haversine pdf}). The PDF describes the likelihood that two meteorites can be found closer than a distance $r$ from each other. The value of the PDF at ${d=10}$~km is 0.0004 in our rectangular working region. Thus, the values in Equation~\ref{eq: nmax}, representing the maximum number of meteorites according to $\lambda$, were multiplied by this PDF value in order to estimate the maximum number of accumulated meteorites with mass higher than 10~g to be found in a 10~km radius.

For instance, in an Antarctic environment \citep[$\lambda = 0.064$~kyr$^{-1}$;][]{Bland96}, where ${N_\text{max}\sim 1.9 \cdot 10^6}$, the maximum number of meteorites in a 10~km radius from each other within the working rectangle would be 847. Similarly, the number of meteorites within a radius of 5~km would be 213 and in a distance of 2~km it would be 34. These values correspond to a maximum density of 2.7 meteorites per km$^2$. This value underestimates the actual value found in Antarctic search expeditions. For example, \citet{Joy19} reported an actual recovered meteorite density of 7.1~km$^{-2}$ and \citet{Cassidy92} stated that this value was 5.6~km$^{-2}$. This disagreement may be explained by the surfacing of meteorites in Antarctic soil due to blue ice movements, which was not taken into account in our simple weathering model based on the $\lambda$ factor.

On the other hand, in the west Sahara region the density of meteorites was estimated to be 1 meteorite per km$^2$ \citep{Aboulahris19}. Applying our model, considering a weathering factor in the region of $\lambda \simeq 0.2$~kyr$^{-1}$, similar to that in Roosevelt County \citep{Zolensky90_weathering}, the maximum number of accumulated meteorites in a $\sim700,000$~km$^2$ area is $6.1\cdot 10^5$ (Figure \ref{fig: nmax}). Using the probability estimations of the PDF, as in the Antarctic case above, this would correspond to 271 meteorites within a 10~km$^2$ radius, that is, a maximum density of 0.9 meteorites per km$^2$, which is in fair agreement with the estimates for the Sahara.

Finally, the case of the Benešov meteorites \citep{Spurny14_Benesov} was also treated with the haversine formalism. As the masses of the found meteorites are lower than our 10~g limit, the flux from \citet{Evatt20_flux} cannot be used for an estimation with the models presented in Section~\ref{subsec: models}. Following the data from \citet{Spurny14_Benesov}, we calculated the probability of finding a random-located meteorite within a 250~m radius in the Czech Republic, for which we assumed $\lambda \sim 1500$~kyr$^{-1}$ (Section~\ref{subsec: lambda in urban regions}). The haversine formalism applied to this case returns a number of meteorites within $d=250~m$ of $N(d) < 1$, which rejects any possiblity that the Benešov meteorites were unrelated. A summary of these calculations can be found in Table~\ref{tab: number density}.

\begin{table}
\caption{Number density of meteorites ($n$) with $m > 10$~g based on the haversine formalism. The haversine PDF curve is derived in Appendix~\ref{app: haversine pdf}. The maximum number of meteorites ($N_\text{max}$) in the working area of 700,000~km$^2$ is calculated from the weathering factor ($\lambda$) using Equation~\ref{eq: nmax} multiplied by the pairing factor $p = 3.18$ \citep{Evatt20_flux}. The number of meteorites in a radius $r$ is depicted in the $N(r)$ column. In the blank cells of the table, read the value in the previous row.}
\label{tab: number density}
\resizebox{\columnwidth}{!}{
\begin{tabular}{lccccr}
\toprule
$\boldsymbol{r}$ \textbf{(km)} & \textbf{Haversine PDF value} & $\boldsymbol{\lambda}  \textbf{ (kyr}\boldsymbol{^{-1}}\textbf{)}$ & $\boldsymbol{N_\text{max}}$        & $\boldsymbol{N\text{\textbf{(}}d\text{\textbf{)}}}$ & $\boldsymbol{n}$ \\ \midrule
10     & $4.4 \cdot 10^{-4}$ & 0.2               & 6.1 $\cdot 10^5$ & 271  & 0.9 \\
       &                     & 0.064             & 1.9$\cdot 10^6$  & 847  & 2.7 \\
5      & $1.1 \cdot 10^{-4}$ &                   &                  & 213  & 0.9 \\
2      & $1.8 \cdot 10^{-5}$ &                   &                  & 34   &    \\ 
0.250  & $2.9 \cdot 10^{-7}$ & 1500              & 81               & <1   & 0.0075 \\ \bottomrule
\end{tabular}
}
\end{table}

It should be noted that the $N_\text{max}$ values used in this work correspond to the maximum number of meteorites to be found in the region (Equation~\ref{eq: nmax}) and multiplying the flux in Equation~\ref{eq: flux conversion} by the pairing factor $p = 3.18$ \citep{Evatt20_flux}, as in this case we modeled the actual number of meteorites. In the other sections of this work, as stated in Section~\ref{subsec: historical}, for simplicity we have considered that each fall generates a single meteorite, and any fragmentation scenario approach was to consider both the terminal and 10-g mass in the ellipse definitions. Moreover, the estimation of $N_\text{max}$ depends on the limiting mass used, which is $m = 10$~g throughout this work. Other number estimations could be obtained by changing this limiting mass.

As a final point, to estimate the likelihood that two such point-modeled meteorites have fallen in close proximity by chance, we can also employ the haversine formula. Based on the data presented in Table~\ref{tab: number density} and using the haversine metric, the probability that two unrelated random points lie within a distance of $2 \cdot 2.2 = 4.4$~km---where 2.2~km corresponds to the expected semi-major axis derived from the area and eccentricity shown in Figure~\ref{fig: ellipses data}---is approximately $10^{-3}\%$. It is important to note, however, that some accumulated meteorites may have undergone fragmentation, resulting in strewn fields that could also be approximated as elliptical in shape, and not single points. In such cases, the probability of spatial coincidence would depend on additional parameters beyond those considered in the present analysis. A comprehensive analysis of this alternative scenario is left for future work.

\subsection{Calculating $\lambda$ in non-desertic regions}\label{subsec: lambda in urban regions}

It is not straightforward to estimate the value of the weathering constant in non-desertic regions. To roughly evaluate the order of magnitude of $\lambda$ in such locations, we performed a simple estimation based on the data of recovered meteorites reported in the Meteoritical Bulletin\footnote{\url{https://www.lpi.usra.edu/meteor/metbull.php}}. For this purpose, we solved for $\lambda$ in Equation~\ref{eq: nmax} by considering a reference area, $A$. Then, $\lambda$ was calculated in terms of the flux of meteorites, $\mathcal{F}(m)$, and the maximum number density of accumulated meteorites, $n_a$, found in that area as
\begin{equation}
    \lambda = \frac{\mathcal{F}(m) \cdot \gamma (\theta)}{n_a}.
    \label{eq: lambda non desertic}
\end{equation}
In order to estimate $n_a$, we used data on reported meteorite finds from the literature. As a matter of example, an upper limit for the particular case of a relatively small country like Austria can be obtained as follows: by October 2024, only three \textit{finds} (i.e., accumulated meteorites following the nomenclature of the present work) from Austria can be found in the Meteoritical Bulletin. The oldest one is the Mühlau ordinary chondrite, found in 1877, while the most recent find corresponds to the Ybbsitz meteorite, classified as an H4 ordinary chondrite. Given that the first find reported in Austria dates back to the 19th century, it can be roughly assumed that the totality of the area of Austria (83,800~km$^2$) has been explored to find accumulated meteorites. This allows extracting an average density of ${n_a = 3.6\cdot 10^{-5}}$ meteorites per km$^2$ in that region. This value must be considered just as an educated guess of the lower limit for the maximum density of meteorites in a region with a temperate, humid climate like Austria.

By inserting this $n_a$ value in Equation~\ref{eq: lambda non desertic} together with the scale factor ${\gamma ( \theta \simeq 40^\circ ) \sim 0.8}$ and the mass flux for ${m>10}$~g (Section~\ref{subsec: historical}), a value $\lambda \sim 1500$~kyr$^{-1}$ is obtained. This is clearly an extreme limit estimation for non-desertic regions, since its calculation is based on highly rough assumptions. For instance, the general population is not trained to look for meteorites, and clearly the whole area of the country has not been exhaustively explored to find meteorites. Also, it cannot be ruled out that there might be casual \textit{finds} that have not been reported. Overall, these factors decrease the number of logged finds in the Meteoritical Bulletin. In any case, the purpose of this estimation is to convey the idea that in non-desertic regions the weathering constant could even be 4 orders of magnitude higher than that in Antarctic regions.

\subsection{Probability estimations from the Poisson distribution approach} \label{subsec: results statistical analysis}

A simple, straightforward approach to calculate the probability of finding an accumulated meteorite inside a fresh meteorite strewn field is to model this process with a Poisson probability distribution (Equation~\ref{eq: Poisson Nh}, Section \ref{subsec: poisson}). This model allowed us to estimate probabilities as a function of the weathering factor, the age of accumulated meteorites on the surface, and the fresh meteorite strewn field areas. Probabilities were calculated considering any coincidence between meteorites of masses higher than 10~g.

\begin{figure*}
    \centering
    \includegraphics[width=1.\textwidth]{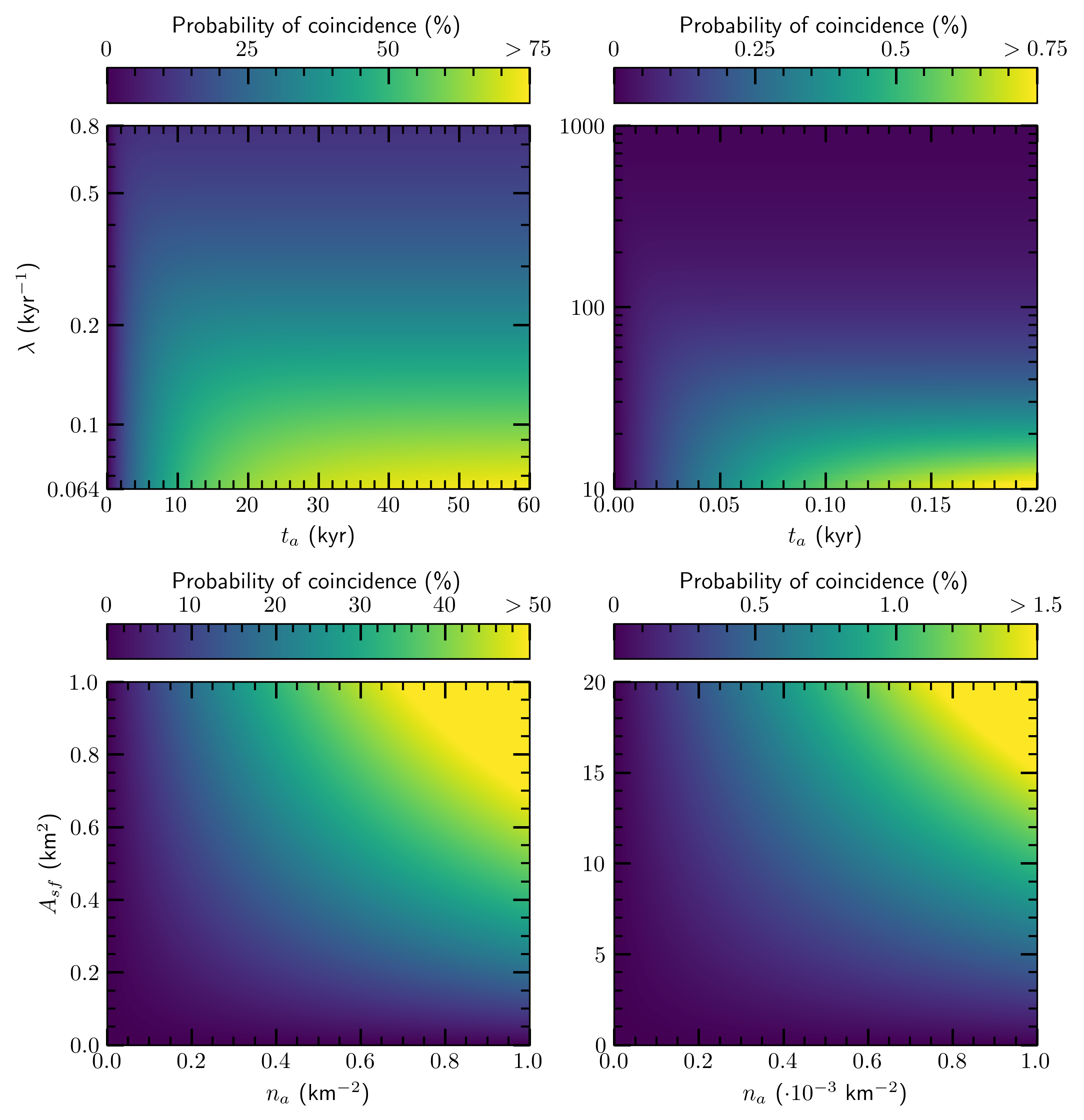}
    \caption{Probability of coincidence of an accumulated meteorite within the strewn field of a fresh meteorite as predicted with the Poisson distribution method. The plots in the left column comprise a set of parameters typical of desertic regions, while the right column represents values associated with non-desertic regions. All fluxes were estimated for a mean latitudinal factor of $\gamma (\bar{\theta}) = 0.801$, corresponding to the Central European region, and with a mass flux of ${\mathfrak{F}(m = 10\;\text{g}) = 68.39}$~km$^{-2}$~Myr$^{-1}$ (Equation~\ref{eq: flux conversion}). \textit{Top:} Dependence with the weathering factor of the environment ($\lambda$) and the terrestrial residence time of accumulated meteorites ($t_a$). Probabilities were calculated with a strewn field area of $A_{sf}= 2.13$~km$^2$, corresponding to the mode of the 1.5 IQR area distribution considering all simulated heights, velocities and masses (Table~\ref{tab: areas summary}). \textit{Bottom:} Dependence with the area of the strewn field ($A_{sf}$) and the number density of accumulated meteorites ($n_a$). The number density of accumulated meteorites includes all the possible $n_a$ values computed with the different $\lambda$ and $t_a$ values in the top row plots.}
    \label{fig: simulations planes}
\end{figure*}

The first row of Figure \ref{fig: simulations planes} shows the results of the dependence of the probability on $\lambda$ and $t_a$. The $\lambda - t_a$ probability plot was calculated with a fixed area of $A_{sf} = 2.13$~km$^2$, which is the mode of the 1.5 IQR strewn field distribution that include all simulated heights, velocities and masses at a ground level of 440~m (Table~\ref{tab: areas summary}). The two plots correspond to a high-probability regime (low $\lambda$ and high $t_a$) and a low-probability regime (high $\lambda$ and low $t_a$). In the top-left plot, the minimum value of $\lambda$ corresponds to Antarctic weathering constant estimation from \citet{Bland96}. Given that no works estimating $\lambda$ can be found in the literature, the upper limit of this parameter in the top-right plot was chosen to consider the results of Section~\ref{subsec: lambda in urban regions}. Note that the selected range also included the value $\lambda = 0.197$~kyr$^{-1}$, which is the estimated value for the Roosevelt County meteorite accumulation site \citep{Zolensky90_weathering}. The maximum value of $t_a$ was chosen to include the stationary state for the lowest possible value of $\lambda$ (Figure~\ref{fig: nmax}). In fact, in both $\lambda - t_a$ probability plots it can be seen that, for each $\lambda$ value, there is a minimum time from which the calculated probability remains constant. In such situations, the system has reached the stationary state and the number of accumulated meteorites is constant regardless of $t_a$ (Figure~\ref{fig: nnmax}), implying a constant coincidence probability.

Regions with low $\lambda$ do properly preserve meteorites for a long time on their surfaces. This is the case for Antarctic regions, for example. Thus, $n_a$ can increase substantially with $t_a$ before reaching the stationary state, yielding a large probability of coincidence. In such cases like Antarctic environments, the probability to find an accumulated meteorite within the strewn field of a fresh meteorite fall can be above 75\%. In an actual scenario, it must be noted that in these highly preservative regions it is often not feasible to compute an actual strewn field of the fall due to the lack of surveillance cameras.

The second row of Figure \ref{fig: simulations planes} shows the results of the dependence of the probability on the strewn field area and the number density of accumulated meteorites. The latter was estimated from $\lambda$ and $t_a$ using Equation \ref{eq: number meteorites} and dividing it by the considered area. The bottom-left plot exemplifies how accumulated meteorite density is the main factor when computing the probability of coincidence, as large probabilities can be obtained even with small swept areas. A region with a high number density of accumulated meteorites would typically be a region with low $\lambda$ but also with meteorites that remained there for long times (i.e., high $t_a$).

In regions such as the Sahara Desert, where the estimated number density of recoverable meteorites is around 1~km$^{-2}$ \citep{Aboulahris19}, the resulting probability of coincidence can reach the 50\% if the considered area is around 1~km$^2$. Again, mass is an important issue. We expect lower meteorite densities as we consider higher masses thresholds, which would decrease the probability of coincidence.

The right column of Figure~\ref{fig: simulations planes} expands the $\lambda$-$t_a$ and $A_{sf}$-$n_a$ probability plots by considering typical scenarios for non-desertic regions. In the $\lambda - t_a$ plot, we modeled a wide $\lambda$ range, as this value is not reported in literature in such environments. As in the left $\lambda - t_a$ plot, the maximum $t_a$ value was chosen so that the stationary state was reached for the minimum possible $\lambda$ value. In general, it can be seen that, for a fixed area of $A_{sf} = 2.13$~km$^2$, the probability of coincidence for non-desertic regions is usually $<0.75\%$. These low values indicate that finding a meteorite from a previous fall in the predicted strewn field is unlikely, but not fully negligible.

We also expanded the calculation of the coincidence probability in a low accumulated meteorite density region and in broader areas regimes (bottom-right plot in Figure~\ref{fig: simulations planes}). The upper density limit in the bottom-right plot was chosen as a representative density of a non-desertic region ($\lambda = 1000$~kyr$^{-1}$ and $t_a = 0.2$~kyr). Even though in Table~\ref{tab: areas summary} we obtained strewn field areas around 2~km$^2$, in reality wider regions are typically swept. Aiming to represent these situations, we took into account areas up to 20~km$^2$. The estimations for such cases show that the coincidence probability is, in general, small. When the expected density of accumulated meteorites is so low (representative of non-desertic regions), the calculated coincidence probabilities are generally lower than 1.5\%, even for searching areas spanning several square kilometers.

\subsection{Probability estimations from Monte Carlo simulations} \label{subsec: results simulations}

Monte Carlo simulations allowed us to validate the results obtained with the statistical approach in Section \ref{subsec: results statistical analysis}. On the one hand, we analyzed the results based on the different occupation numbers of the ellipses as a function of $\lambda$ and $t_a$ (Section \ref{subsec: lambda t indiv}). The results thus obtained are compared with those of the Poisson distribution method (Section \ref{subsec: lambda th plane}). For consistency with the Poisson analysis, the simulations included the ellipses estimated from all the different combinations of initial heights, velocities and masses, at 440 m ground level (red plot in Figure~\ref{fig: ellipses data}). However, the results of this section would be extrapolable to any specific case of dark flight initial conditions.

\subsubsection{Dependence on $\lambda$ and $t_a$} \label{subsec: lambda t indiv}

\begin{figure*}
    \centering
    \includegraphics[width=1.\textwidth]{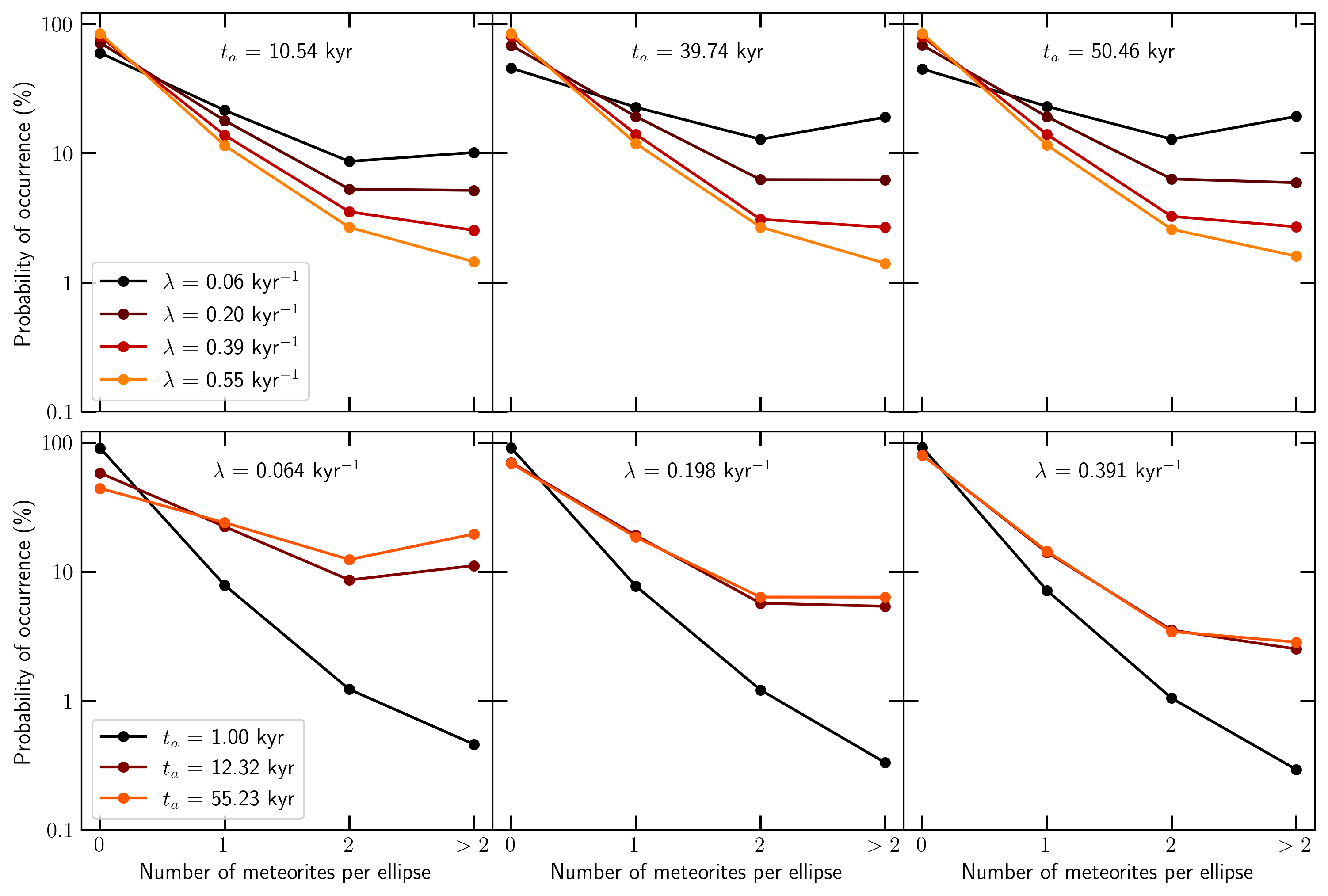}
    \caption{Probability of occupation of the strewn fields for different weathering constants ($\lambda$) and integration times ($t_a$), as computed from Monte Carlo simulations. Results are presented for fixed $t_a$ (first row) and for fixed $\lambda$ (second row). The different $\lambda$ and $t_a$ values chosen span the regimes for meteorite accumulation (Figure \ref{fig: nnmax}).}
    \label{fig: individual lambda th}
\end{figure*}

Figure \ref{fig: individual lambda th} shows individual results for coincidences at fixed $\lambda$ and $t_a$ as obtained with Monte Carlo simulations. The values of $\lambda$ and $t_a$ in Figure \ref{fig: individual lambda th} were chosen as representative of different regimes of meteorite accumulation (Figure \ref{fig: nnmax}). The values $\lambda = 0.064$~kyr$^{-1}$ and $\lambda = 0.198$~kyr$^{-1}$ represent the Antarctic and Roosevelt County cases, respectively \citep{Bland96, Zolensky90_weathering}. The value $\lambda = 0.391$~kyr$^{-1}$, extracted from the grid of simulated $\lambda$ values (Section~\ref{subsubsec: montecarlo}), was intended to represent a higher weathering region. The different values of $t_a$ span the regime where the number of accumulated meteorites is still increasing and also the case where the stationary state has been reached (Figure~\ref{fig: nnmax}). Probabilities are given in terms of the occupation numbers of the ellipses (one or more meteorites per ellipse). 

Both $\lambda$ and $t_a$ determined the final number of accumulated meteorites generated with Monte Carlo simulations (Figure~\ref{fig: nnmax}). As such, for a fixed value of $t_a$, in a highly weathering environment (higher $\lambda$), the number of coincidences decreased due to the lower number of accumulated meteorites in place. Besides, the probability of finding one or more meteorites per ellipse decreased with increasing $\lambda$ (see lower panels), as fewer accumulated meteorites were present in the working region. On the other hand, if $\lambda$ was fixed, the number of accumulated meteorites was found to increase with increasing $t_a$, since ellipse occupation is more favorable. In turn, the occupation number of the ellipses increased due to the larger availability of accumulated meteorites. This can be noticed in the panels with $\lambda$ fixed, but also when sequentially following the same $\lambda$ line in the $t_a$ row. 

It is also worth noticing that the variation in the occupation ratio was more dependent on $\lambda$ than on $t_a$. This was already predicted by the Poisson distribution method (Figure \ref{fig: simulations planes}). For a fixed $t_a$, the probability of coincidence could change an order of magnitude with $\lambda$. On the other hand, for a fixed $\lambda$, a variation of 40~kyr may not introduce great changes in the probability. This was specially relevant for high $\lambda$ values, as the simulated $t_a$ often exceeded the minimum time required to reach stationary state. In the case of $\lambda = 0.391$~kyr$^{-1}$, for example, such time is $\sim10$~kyr (Figure~\ref{fig: nnmax}). Therefore, the lines at $t_a = 12.32$~kyr and $t_a = 55.23$~kyr both represented the same number of accumulated meteorites, thus showing a similar behavior. Variations between these two lines are due to the statistical behavior of the Monte Carlo simulations.

\subsubsection{The $\lambda - t_a$ probability plot from Monte Carlo simulations} \label{subsec: lambda th plane}

\begin{figure}
    \centering
    \includegraphics[width=\columnwidth]{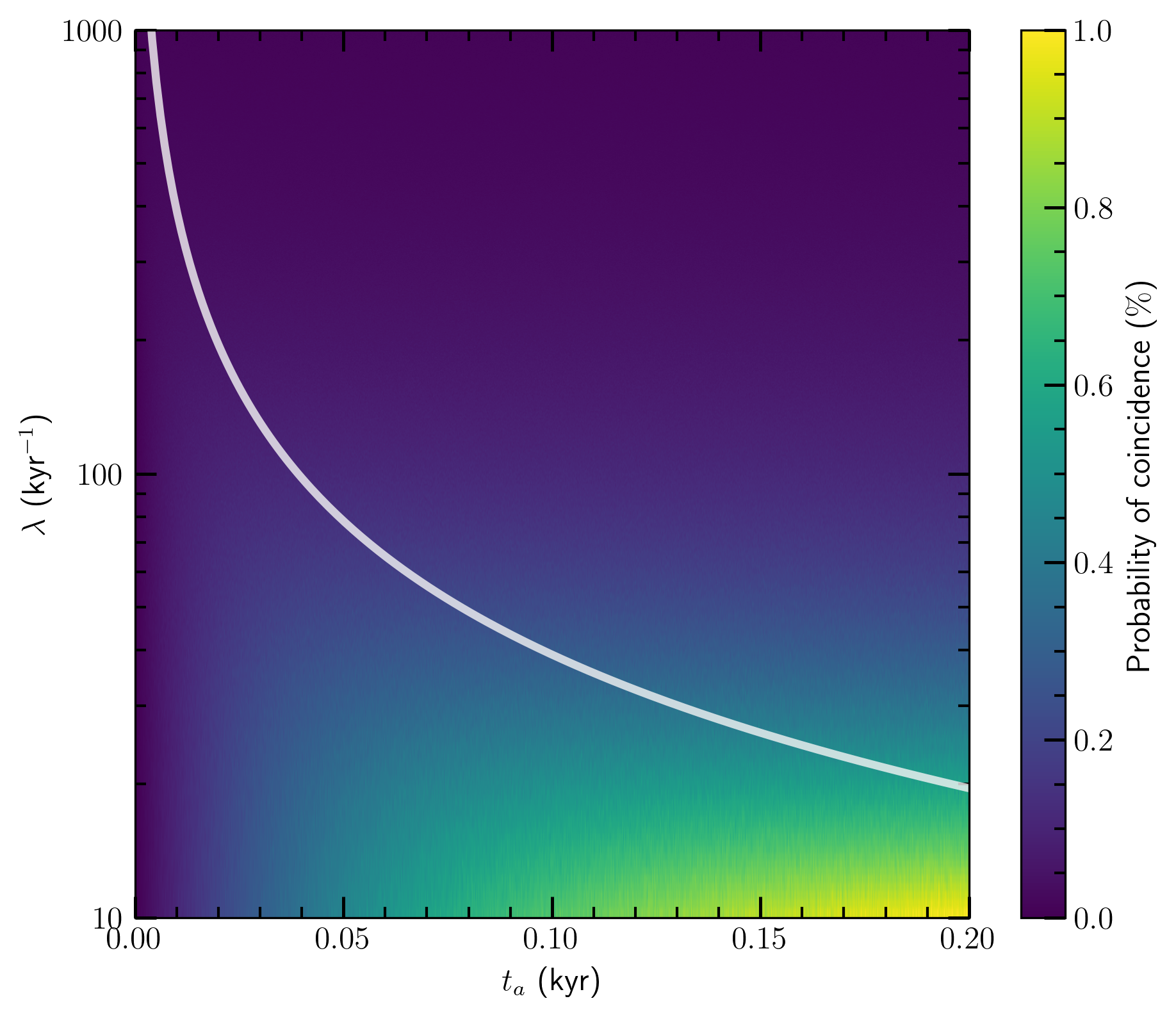}
    \caption{Plot of the probability of coincidence based on the weathering factor ($\lambda$) and the time of integration ($t_a$), as computed from Monte Carlo simulations. The probability is calculated from the number of occupied fresh meteorite ellipses. The value is strongly dependent on the number of accumulated meteorites simulated according to the $(\lambda, t_a)$ pair value. The white line corresponds to the time value required to reach the stationary state, where the number of accumulated meteorites becomes constant for a constant $\lambda$.}
    \label{fig: lambda t plane}
\end{figure}

We combined the results from Section \ref{subsec: lambda t indiv} for different values of $\lambda$ and $t_a$. The result was the $\lambda -t_a$ probability plot shown in Figure \ref{fig: lambda t plane}, where the probability of coincidence is represented. The plot limits were chosen to include all relevant values of $\lambda$ and $t_a$ in a low-coincidence regime, as in the top-right plot in Figure~\ref{fig: simulations planes}. This would be the regime for non-desertic regions (Section~\ref{subsec: lambda in urban regions}). The $\lambda - t_a$ grid included 1000 divisions along each axis. Thus, it included $10^6$ different $(\lambda, t_a)$ pairs coincidence probability, each value estimated from the occupation of $3.92\cdot 10^5$ ellipses (Section~\ref{subsec: fresh}). The probability shown in Figure \ref{fig: lambda t plane} corresponds to the probability that a strewn field may be occupied by any non-zero number of accumulated meteorites, while in Figure~\ref{fig: individual lambda th} probabilities are shown by occupation numbers.

The white line in Figure~\ref{fig: lambda t plane} shows the stationary state time relation, as calculated from Equation \ref{eq: n model} considering a meteorite flux of masses higher than 10~g. We considered that the stationary state was reached when $n_a/n_\text{max} \geq 0.98$. Note that if we chose a higher value of the minimum meteorite mass, then the flux value would be lower, requiring a less amount of time to reach the stationary state, for a given $t_a$.

To the right of this curve, the number of accumulated meteorites was the same for a given $\lambda$ value. This would explain the scarce variation of probability with $t_a$ at high $\lambda$ values seen in Figure \ref{fig: individual lambda th}. Most of the points in the plot lie in the stationary state region. The highest variation is seen at low $\lambda$ values, where there is a higher time span before reaching this state. To the left of the white curve, the number of accumulated meteorites strongly varies with $t_a$, starting at $N_a = 0$.

\begin{figure}
    \centering
    \includegraphics[width=\columnwidth]{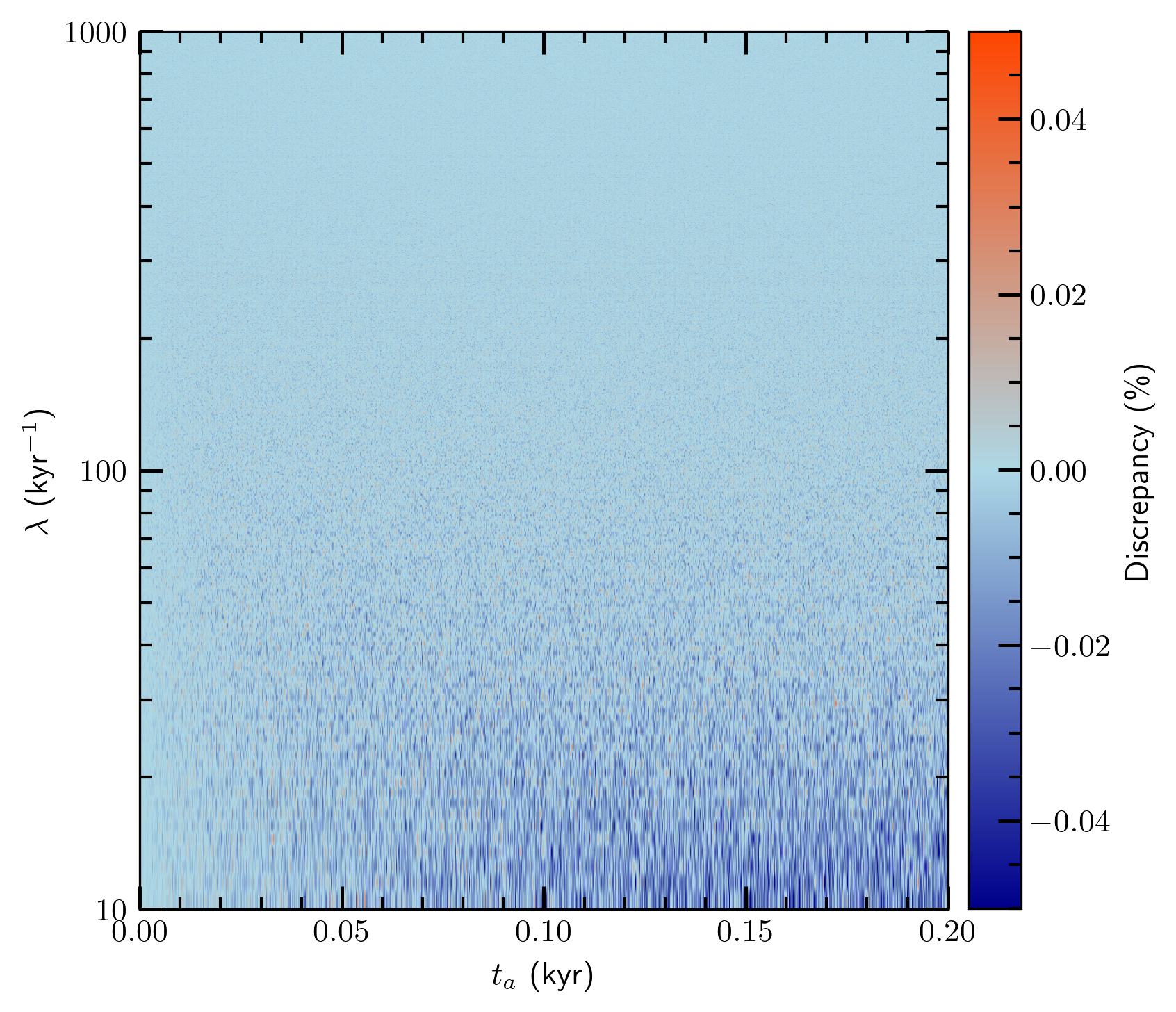}
    \caption{Discrepancy between the probability of coincidence predicted by the Monte Carlo method (Figure~\ref{fig: lambda t plane}) and the Poisson estimation (top-right plot in Figure~\ref{fig: simulations planes}). Plotted values are the subtraction of the Poisson estimation from the Monte Carlo estimation. Blue values indicate regions where the Poisson model predicts a higher probability of coincidence, while orange values correspond to regions where the Monte Carlo method yields higher probabilities. In absolute values, discrepancies are less than 0.05\%, meaning that both models are comparable.}
    \label{fig: model comparison}
\end{figure}

The $\lambda -t_a$ probability plot calculated from a Poisson distribution (Figure~\ref{fig: simulations planes}) remarkably resembles the $\lambda -t_a$ probability plot calculated from simulations (Figure~\ref{fig: lambda t plane}), particularly in the stationary state region. For completeness, Figure~\ref{fig: model comparison} compares the top-right plot of Figure~\ref{fig: simulations planes} with the $\lambda - t_a$ plane in Figure~\ref{fig: lambda t plane}. The plot is obtained by subtracting the Poisson estimation from the Monte Carlo estimation, value-by-value. Overall, we find that the Poisson method overestimates the coincidence probability by only 0.05\% compared to the Monte Carlo approach. The subtle differences are due to the statistical behavior of the Monte Carlo method, driven by the non-constant area of the ellipses. Therefore, our model (Equation~\ref{eq: Poisson Nh}) provides a suitable fast-calculating approach to estimate the probability that a non-related meteorite can be found within the strewn field of a witnessed fall.

\subsection{Application to real cases} \label{subsec: application to real cases}
The Lake Frome 006 meteorite was a 30~g meteorite found in a hot desert with a terrestrial age of $t_a = 3.2 \pm 1.3$~kyr \citep{Devillepoix22_minimoon}, after the research team swept an area of 0.7~km$^2$. For the probability of coincidence in this case, we assumed a weathering constant of $\lambda \sim0.2$~kyr$^{-1}$, similar to that from Roosevelt County \citep{Zolensky90_weathering}, a flux of incoming meteorites of $\mathfrak{F}(m = 30\;\text{g}) = 45.7$~km$^{-2}$~Myr$^{-1}$ and a latitudinal factor of $\gamma (\theta) = 0.940$ \citep{Evatt20_flux}. With these assumptions, the probability of finding the Lake Frome 006 meteorite in a computed fresh strewn field is $\sim7\%$, as estimated with Equation \ref{eq: Poisson Nh}. In an alternative estimation, \citet{Devillepoix22_minimoon} showed that the upper limit of this probability would be $2\%$, which is quite lower than our estimation. In order to get their values for probability, our Poisson model would require $\lambda \sim 1.5$~kyr$^{-1}$. 

On the other hand, the Ischgl meteorite \citep{Gritsevich24_Ischgl} was found within the computed strewn field of the EN241170 fireball, in Austria, a highly weathering Alpine environment. In these regions, meteorites are expected to have relatively short preservation times due to erosion and burial processes, thereby reducing the probability of coincidental discovery. For Ischgl, the lack of \ce{^60 Co} suggested this meteorite was fresh. As mentioned in Section~\ref{subsec: historical}, we assumed that a meteorite is not fresh anymore when its residence time exceeds 10 years. We used the Poisson distribution expression (Equation~\ref{eq: Poisson Nh}) to assess the probability that a meteorite can land in a fresh meteorite strewn field within the timescale of 10 years. We considered a weathering factor of $\lambda = 1500$~kyr$^{-1}$, which was calculated from the methodology in Section~\ref{subsec: lambda in urban regions}. Additionally, we assumed a flux of ${\mathfrak{F}(m = 1000\;\text{g}) = 5.8}$~km$^{-2}$~Myr$^{-1}$, as the mass of the Iscghl meteorite was 1~kg, and a latitudinal factor of $\gamma (\theta ) = 0.790$. The strewn field area for the computed meteoroid probably associated with the Iscghl meteorite was 210~km$^2$. Under these conditions, the probability that another meteorite may have landed on the same region of the computed strewn field was 0.06\%, as estimated with the Poisson distribution method (Equation~\ref{eq: Poisson Nh}).

\begin{figure}
    \centering
    \includegraphics[width=\columnwidth]{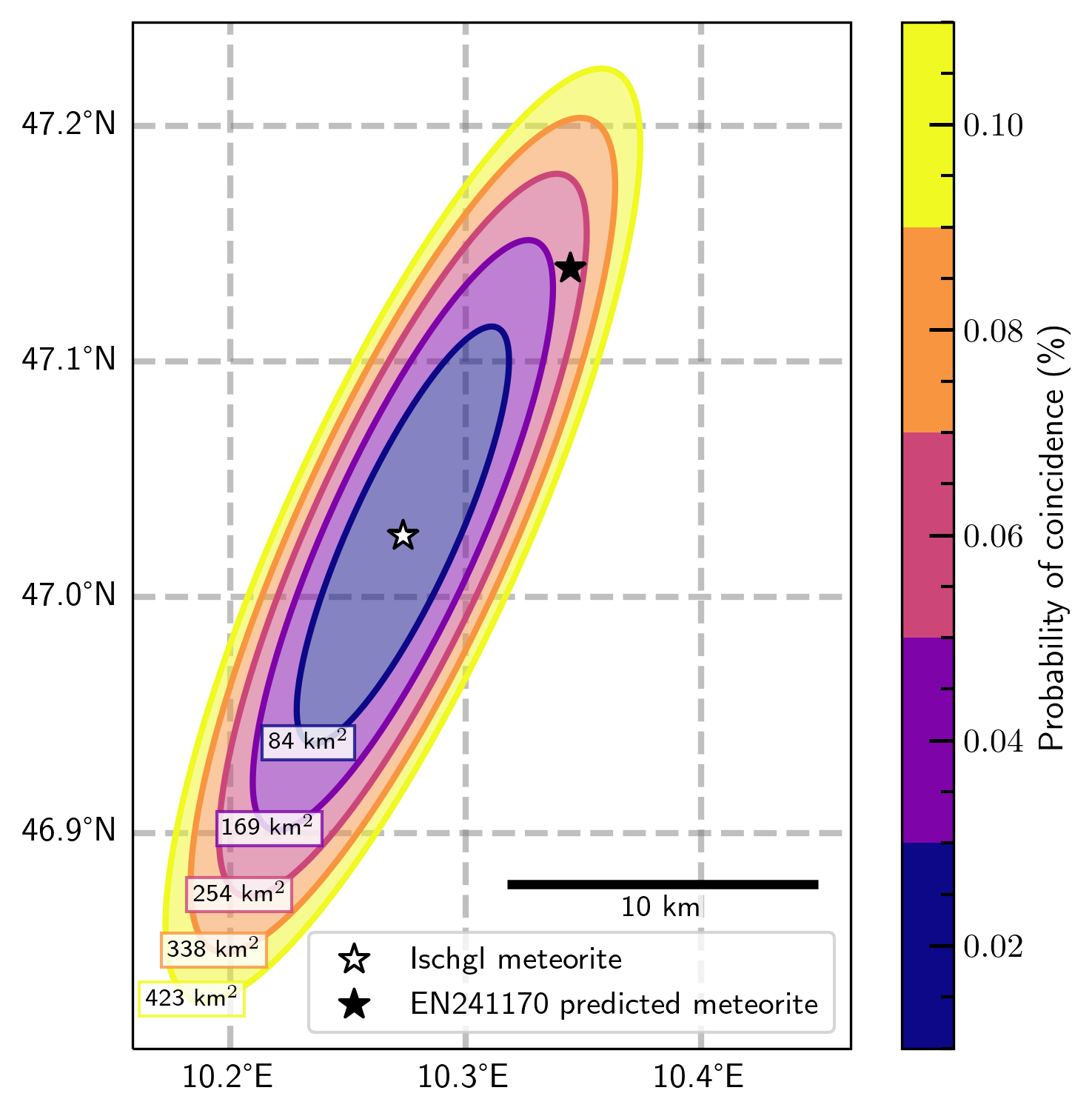}
    \caption{Equiprobability ellipses for coincidence with previous unrelated meteorites for the Ischgl meteorite case ($m = 1$~kg, $\lambda \sim 1500$~kyr$^{-1}$ and $t_a = 10$~yr). The actual Iscghl meteorite finding site is shown at the center of the ellipses as a white star, and the predicted EN241170 meteorite is shown as a black star \citep{Ceplecha1977_EN241170fireball}. The ellipses eccentricity and orientation are extracted from the most recent EN241170 prediction \citep{Gritsevich24_Ischgl}. The area of each ellipse in squared-kilometers is tagged correspondingly. Scale bar is 10~km. The overall low probability of coincidence supports that the Ischgl meteorite most likely originated from the EN241170 fireball.}
    \label{fig: heatmap}
\end{figure}

As a matter of example, Figure~\ref{fig: heatmap} shows a heatmap distribution of the probability of finding an unrelated meteorite around the actual Ischgl finding site. The eccentricity of the ellipses and its azimuth were extracted from \citet{Gritsevich24_Ischgl} dark flight estimations. Additionally, the former prediction of the EN241170 meteorite fall location \citep{Ceplecha1977_EN241170fireball} is shown. Even though the Ischgl fireball strewn field was 210~km$^2$, the area within the strewn field where 1~kg fragments can be found with, at least, 0.02\% of probability is much smaller ($\sim 80$~km$^2$). Overall, the estimated values show a fairly negligible probability of coincidence that the Ischgl meteorite was unrelated to the EN241170 fireball, thus supporting the conclusions in \citet{Gritsevich24_Ischgl}.

Another case in which our methodology can be applied is the Almahata Sitta meteorite. When it fell in 2008, it was straightforwardly related to the asteroid 2008 TC$_3$ \citep{Jenniskens09_AlmahataSitta}, which was classified as a polymict chondrite breccia that disrupted in the atmosphere \citep{Bischoff22_AlmahataSitta}. During 3 months after the observed fall, a research team swept an area of $\sim 70$~km$^{2}$ in the Nubian Desert. The Meteoritical Bulletin entry for Almahata Sitta includes 57 strewn field members found within these 3 months, including 45 ureilites, 5 ordinary chondrites and 7 unclassified samples. Assuming $\lambda \sim 0.2$~kyr$^{-1}$ for the Sahara Desert, we can apply Equation~\ref{eq: Poisson Nh} to this case. For a minimum mass of ${m=10}$~g, the lower mass limit by \citet{Evatt20_flux}, the probability of having a non-related meteorite in that area over 90 days is $11.3\%$. This non-negligible probability might indicate that some of the recovered meteorites from the Almahata Sitta strewn field were already there prior to the 2008 TC$_3$ fall. Such coincidence would support the variety of CREA seen among the Almahata Sitta stones \citep{Goodrichetal_2019}.

Finally, we ought to underline that we did not apply our model to estimate the probability of coincidence for the Raja fall, as the mass of the recovered old meteorite (3.70 g) is below the lower mass threshold of the baseline dataset used in this study.

These results emphasize the relevance of isotopic dating in meteorites from presumable falls under study \citep{Gritsevich24_Ischgl}. Only really fresh meteorites might contain SLNs due to their recent exposure to the interplanetary medium.

\section{Summary and final remarks}\label{sec: conclusions}
In this work, we made the first systematic approach to model the probability that a previously fallen meteorite can be found in the strewn field of a fresh meteorite. This estimation was assessed from two different points of view. On the one hand, the probability was evaluated considering that the process can be described by a Poisson probability distribution. This method was used to characterize the probability of coincidence as a function of the number density of accumulated meteorites inside a given strewn field area. The Poisson distribution method is shown to be particularly useful to model this kind of problems, requiring as inputs the weathering constant of the environment, the terrestrial age of the accumulated meteorites and the fresh meteorite strewn field area. Acting as a validity test, we performed additional calculations of the probability of coincidence by means of Monte Carlo simulations. The strewn field area distribution was computed based on real fireball data.

Both types of simulations required us to first estimate the number of accumulated meteorites. This was done with a simple model that accounts for both meteorite flux income and terrestrial weathering, assuming that each fireball generated a single meteorite. For simplicity, terrestrial weathering was the only considered meteorite-loss factor. Based on previous works, this factor was assumed to obey an exponential law. We also assumed that it did not depend on the meteorite mass. This approach should be improved in further works, as one may expect that smaller meteorites are more prone to be lost by weathering than bigger meteorites, due to their higher environmental interaction for their larger surface area to volume ratio. Additionally, accounting for the fragmentation of accumulated meteorites could lead to an increased number of meteorites reaching the ground, each with correspondingly lower individual mass. While this might elevate the probability of spatial or temporal coincidence, the expected dispersed distribution pattern on the ground would likely mitigate such effects. In the absence of detailed fragmentation modeling, the probabilities presented in this study might be regarded as conservative lower estimates.

Considering the general case of simulated fresh meteorites, their landing region was computed from a dark flight simulation based on real fireball data from the EN. Such general estimations could be further improved with an expanded dataset. The region where a research team would go searching for a fresh meteorite was estimated from an ellipse, calculated from the landing points from the dark flight. This methodology resembles that of a real case. The ellipses were generated from different initial heights, velocities and masses, the areas of which are summarized in Table~\ref{tab: areas summary}. With this, we expected to include all possible ellipse area ranges.

Table~\ref{tab: summary probabilities} shows a summary of different probabilities calculated with the Poisson distribution method as applied to different, real cases. In the table, the limiting mass for each case, which allowed us to calculate the corresponding mass fluxes, is also indicated. For a particular set of $\lambda$ and $t_a$ values, which allowed us to compute the accumulated meteorite density, together with a search area value, the coincidence probability was estimated. The latter is given in the last column of the table, as calculated with the Poisson distribution method.

\begin{table*}
\centering
\caption{Summary of the probability that an accumulated meteorite more massive than a given mass can be found inside a swept strewn field of area $A_{sf}$ of a fresh meteorite. The meteorite flux given is the equatorial flux and depends on the minimum mass considered. The $\gamma (\theta)$ factor scales the meteorite flux based on the region latitude. Probabilities were calculated with the Poisson distribution method, starting from an assumption of the weathering factor ($\lambda$) and terrestrial age of the accumulated meteorites ($t_a$), from which the number density of accumulated meteorites ($n_a$) was calculated. By setting $t_a = +\infty$ we assured that the probability was computed in the stationary state. Antarctic cases are additional illustrative examples of the method not included in the main text.}
\label{tab: summary probabilities}
\resizebox{\textwidth}{!}{
\begin{tabular}{lcccccccr}
\toprule
\textbf{Case} & \textbf{Min. mass (g)} & \textbf{Flux (km}$\boldsymbol{^{-2}}$\textbf{ Myr}$\boldsymbol{^{-1}}\textbf{)}$  & $\boldsymbol{\gamma}\textbf{(}\boldsymbol{\theta}\textbf{)}$ & $\boldsymbol{t_a}$ \textbf{(kyr)} & $\boldsymbol{\lambda}$  \textbf{(kyr}$\boldsymbol{^{-1}}$\textbf{)} & $\boldsymbol{n_a}$ \textbf{(km}$\boldsymbol{^{-2}}$\textbf{)} & $\boldsymbol{A_{sf}}$ \textbf{(km}$\boldsymbol{^2}$\textbf{)} & \textbf{Probability (\%)}\\ \midrule
Antarctic & 10 & 68.4 & 0.61 & $+\infty$ & 0.064 & 1.1 & 2.13 & 75     \\
          &       &   &      &         &       &     & 10   & 99.8            \\
          & 30 & 45.7 &      &         &       & 0.7 & 2.13 & 60.5            \\
          & 50 & 37.8 &      &         &       & 0.6 &      & 53.6            \\
          & 100 & 29.3 &     &          &      & 0.4 &      & 44.8            \\ 
\vspace{1em} & & & & & & & & \\
Lake Frome 006 & 30 & 45.7 & 0.94 & 3.2 & 0.2 & 0.1  & 0.7 & 6.9 \\
               &    &      &      &   & 1.5 & 0.03 &     & 2.0             \\
\vspace{1em} & & & & & & & & \\
Ischgl & 1000 & 5.8 & 0.79 & 0.01 & 1500 & $6\cdot 10^{-6}$ & 210 & 0.06  \\
\vspace{1em} & & & & & & & & \\
Almahata Sitta & 10 & 68.4 & 0.96 & $2.5 \cdot 10^{-4}$ & 0.2 & $1.6\cdot 10 ^{-5}$ & 70 & 11.3 \\\bottomrule

\end{tabular}
}
\end{table*}

Next, the main results and remarks derived from this work are summarized:

\begin{itemize}
    \item Meteorite accumulation in a given site may reach a constant number after very long times, typically of the order of thousands of years. This depends on the environment conditions, which can be characterized by a weathering factor $\lambda$. The number of non-fresh (accumulated) meteorites in a given area can be reset after natural or anthropogenic events. Such resetting events are typical of more populated areas.  
    \item The Poisson approach was proven a fast-computing, effective model to estimate the probability of coincidence between accumulated and fresh meteorites. In order to compute the probability, only a weathering constant, a residence time for accumulated meteorites and the area of the strewn field are required. The method was successfully validated using Monte Carlo simulations.
    \item According to the Poisson modeling, the shape of the strewn field is not relevant for probability estimation, but only its area.
    \item The haversine expression can be used to estimate the number density of meteorites within a fixed region on Earth. In regions with Antarctic climatology, the number of meteorites from different falls in a 10~km radius was estimated to be $\sim840$.
    \item From reported \textit{finds} data on the Meteoritical Bulletin, we estimated that the weathering constant in non-desertic regions can be 4 times higher than those in Antarctic regions.
    \item For meteorites with masses larger than 10~g and strewn field areas of ${A_{sf} = 2.13}$~km$^2$, in environments with low $\lambda$, such as Antarctica, the probability can be higher than 75\% as old as than ${\sim 20}$~kyr. In regions with hot desert climatology, this probability drops to $\sim 45$\%. This effect is due to the highly meteorite-preserving conditions of these regions.
    \item In highly weathering environments where the number density of accumulated meteorites with masses larger than 10~g is low, such as the countryside, the probability of coincidences is of the order of $< 1\%$. Even though having a low coincidence probability, these results suggest that meteorite samples found inside such areas, should have their terrestrial age characterized with its SLNs abundances before relating it with a recorded fall.
    \item According to the Poisson distribution modeling, for ${\lambda \sim 0.2}$~kyr$^{-1}$ the probability of finding the Lake Frome 006 meteorite within a fresh strewn field of 0.7~km$^2$ is $\sim 7$\%, considering meteorite masses larger than 30~g. Thus, a coincidence in a desert is not completely unlikely accounting for its low-weathering environment. In order to have a probability about 2\% \citep{Devillepoix22_minimoon} the weathering factor should have been $\lambda \sim 1.5$~kyr$^{-1}$.
    \item The probability of coincidence for the Ischgl meteorite fall is $0.06\%$, considering a weathering constant of $\lambda \sim 1500$~kyr$^{-1}$ for Austria, meteorite masses larger than 1~kg, and a strewn field area of $A_{sf} = 210$~km$^2$. This low probability value reinforces the conclusions of \citet{Gritsevich24_Ischgl}, for which the EN241170 fireball would be the source of the Ischgl meteorite, additionally supported by the W0 weathering degree of this meteorite.
    \item In the case of Almahatta Sitta, our model suggests a probability of coincidence with an unrelated fall of 11.3\% within a strewn field area of 70~km$^2$. This prediction could explain the differences seen between the CREA of different Almahata Sitta fragments and reinforce that meteorite pairing should be conducted carefully.
\end{itemize}

As a final remark, we ought to underline that our results strongly depended on the number of accumulated meteorites considered. All our estimations relied on values taken from scientific publications, and no numerical assumptions were made. On the other hand, in our model, we assumed terrestrial weathering as the only meteorite-loss factor. Other causes may be meteorite burial, accidental displacements by animals, other natural phenomena, or the own ability of humans to find and remove meteorites from the strewn fields. Overall, the present results provide the coincidence probability in desertic regions and the coincidence probability lower limit for non-desertic regions. While our $\lambda$ estimations aim to include these effect as well, its values are an approximation of a more complex process. Isotopic analysis of freshly recovered meteorites should be strongly considered before stating any association with a witnessed fall, as the spatiotemporal coincidence of accumulated and fresh meteorites is non-negligible even in non-desertic regions, specially when the explored areas span several square kilometers.

\section*{Acknowledgments}
We sincerely thank M. Gritsevich for the valuable feedback, which significantly improved the quality of this manuscript. Financial support from the project PID2021-128062NB-I00 funded by MCIN/AEI/10.13039/501100011033  is acknowledged, as well as the Spanish program Unidad de Excelencia María de Maeztu CEX2020-001058-M. PG-T acknowledges the financial support from the Spanish MCIU through the FPI predoctoral fellowship PRE2022-104624. EP-A acknowledges financial support from the LUMIO project funded by the Agenzia Spaziale Italiana (2024-6-HH.0). This work is part of the doctoral thesis of PG-T, within the framework of the Doctoral Program in Physics at Universitat Autònoma de Barcelona.

\section*{Author statements}
\textbf{PG-T:} Methodology, Formal analysis, Software, Writing - Original draft preparation. \textbf{EP-A:} Conceptualization, Methodology, Writing - Review \& Editing. \textbf{JMT-R:} Supervision, Writing - Review \& Editing, Funding acquisition. \textbf{JI-I:} Supervision, Writing - Review \& Editing, Funding acquisition.


\section*{Data Availability}
 
All the data used in this work can be found in its related references, particularly in \citet{Evatt20_flux} and in \citet{Borovicka2022_EFNdataI}. Modeled data will be provided upon request.



\bibliographystyle{mnras}
\bibliography{bibliography}




\appendix
\section{Computational estimation of the PDF of the haversine function} \label{app: haversine pdf}

The haversine equation (Equation~\ref{eq: haversine}) has been proven useful for estimating the distance between two points on the surface of a sphere. This estimation provides simple but accurate calculations without the need of solving the geodesics on Earth's surface.

\begin{figure}
    \centering
    \includegraphics[width=1\columnwidth]{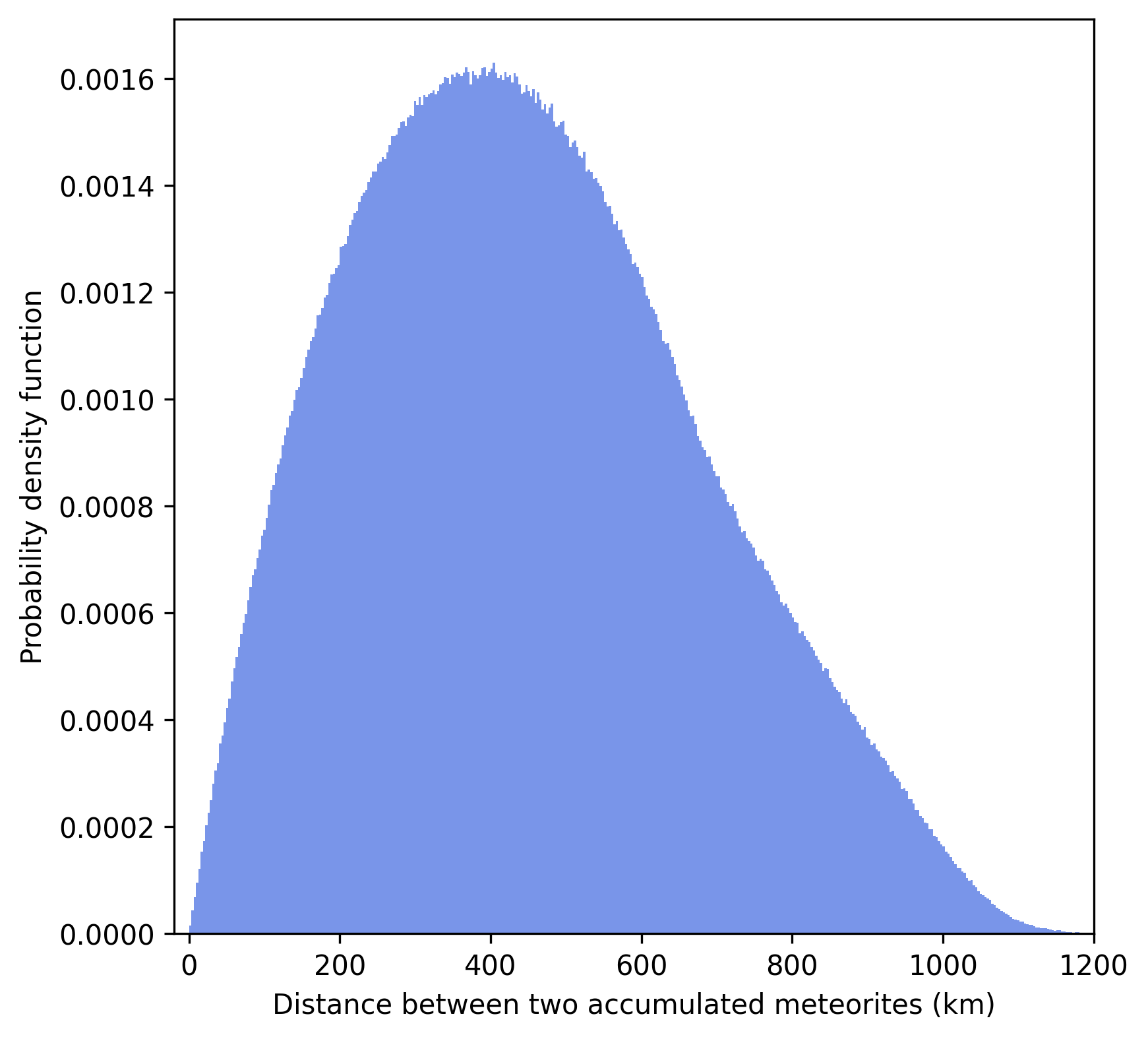}
    \caption{Histogram of the distance of $10^7$ pairs of random points on the working rectangle, with integrated area equal to 1. The histogram binning was chosen with the Freedman Diaconis estimator.}
    \label{fig: haversine pdf}
\end{figure}

As the expression of the haversine involves complex dependencies with the latitude and longitude of the comparing points, we have estimated the PDF computationally. The value of the PDF returns the likelihood that a given pair of points uniformly distributed on a sphere can be closer than a distance $D$. We generated random pairs of points on the working rectangle in Central Europe, defined by the vertexes coordinates ($\theta_\text{min} = 8.745858^\circ \text{ E},\; \phi_\text{min} = 47.447117^\circ \text{ N}$) and ($\theta_\text{max} = 23.552000^\circ \text{ E},\; \phi_\text{max} = 53.4607833^\circ$ N), as denoted in Section~\ref{subsec: fresh}. We computed the distance between a pair of points using the haversine expression (Equation~\ref{eq: haversine}). We evaluated the distances for $10^7$ pairs, from which the normalized histogram, a proxy for the PDF, is shown in Figure~\ref{fig: haversine pdf}. 

Each point was generated from a uniform random distribution on the surface of a sphere. Latitude ($\phi$) and longitude ($\theta$) of each point were taken from the following distributions:
\begin{equation}
\begin{cases}
\theta (u) = 2\pi u,  & {u \in \left[ \frac{\theta_\text{min}}{2\pi},\; \frac{\theta_\text{max}}{2\pi}\right]}\\
\phi (v ) = \arccos\left( 2v -1 \right),  & {v \in \left[ \frac{1 + \cos\phi_\text{min}}{2}, \; \frac{1 + \cos\phi_\text{max}}{2} \right]}
\end{cases},
\end{equation}
where $u, v$ are uniformly distributed variables in the mentioned ranges.

We evaluated the value of the PDF for a distance $D$ by dividing the number of pairs with a distance $r<D$ by the total number of simulated pairs, that is, $10^7$.

\section{The accumulated and recent fluxes of falls} \label{app: mass distribution comparison}

\begin{figure}
    \centering
    \includegraphics[width=1\linewidth]{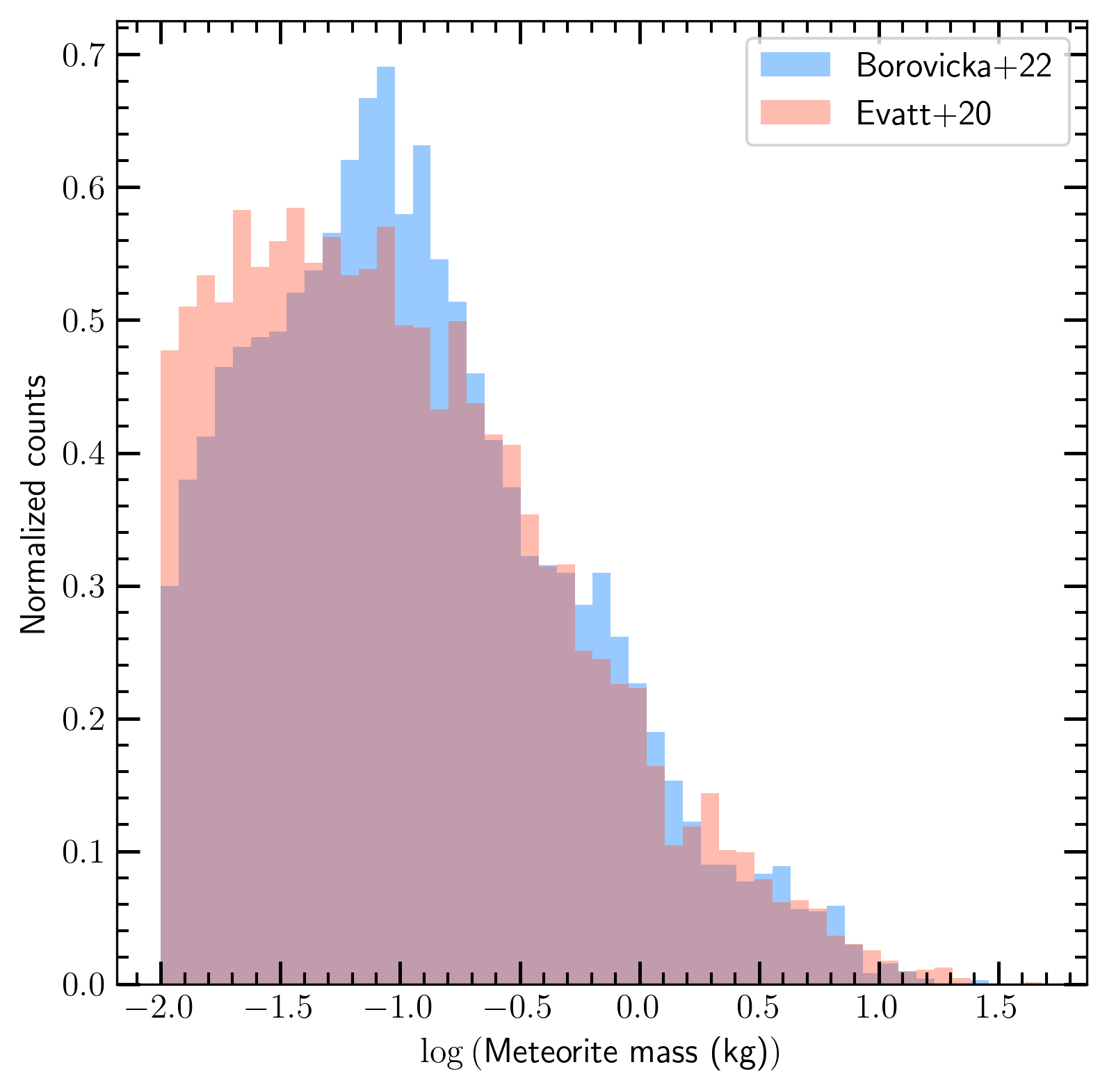}
    \caption{Histograms for 10,000 masses generated with the values from \citet{Borovicka2022_EFNdataI} (recent) and from \citet{Evatt20_flux} (accumulated). Each histogram includes the same 50 bins over the represented range.}
    \label{fig: mass distributions comparisons}
\end{figure}

It is often assumed that the flux of falls has not substantially changed in the last geological epoch. As explained in Section~\ref{subsec: historical}, the incoming flux of meteorites from \citet{Evatt20_flux} was estimated with Antarctic meteorite data. These meteorites had a mean surface residence timescale of $\sim7.2$~kyr. This allowed to estimate the flux over a range of kiloyears. Starting with a cumulative flux distribution, one can derive the PDF of the mass of the incoming meteorites by involving statistical analysis. In such a case, the cumulative flux distribution was used to compute the cumulative density distribution (CDF), from which the mass PDF was derived. Figure~\ref{fig: mass distributions comparisons} presents the distribution of 10,000 masses generated with a PDF computed with this methodology.

On the other hand, in Section \ref{subsec: fresh} we generated new meteoroids with realistic parameters distributions according to the EN data \citep{Borovicka2022_EFNdataI}, which reports events in only 2 years of observation. With the same methodology as described in Section~\ref{subsec: fresh}, masses of incoming meteoroids were also created.

As seen in Figure \ref{fig: mass distributions comparisons}, both mass distributions are nearly identical. In a logarithmic scale, it is seen that the tails at masses higher than $\sim350$~g coincide between both distributions. At lower masses, there is a slight change in the shape of the curve. The distribution from \citet{Borovicka2022_EFNdataI} peaks at $\log m = -1.06$, while the distribution from \citet{Evatt20_flux} peaks at $\log m = -1.28$, which has a higher presence of low-mass meteorites. This could be due to observational effects of the EN cameras, which may be related to night-only observations, cloudy nights or fireball detection. In fact, less massive meteoroids may be related with fainter fireballs, which are often harder to observe.

Nevertheless, both works thoroughly represent the distribution of meteorites at substantially high masses, despite being independent calculations. While one distribution was taken during 2 years of observation, the other includes data of kiloyears. Thus, we consider fair to assume that the incoming flux of meteorites in terms of mass distribution has no relevant changes in the last kiloyears.

\bsp	
\label{lastpage}
\end{document}